\documentclass[draft]{agujournal2019}
\usepackage{url} 
\usepackage[inline]{trackchanges} 
\usepackage{soul}
\usepackage{amsmath}
\usepackage{amssymb}
\usepackage{caption}
\usepackage{subcaption}
\usepackage{siunitx} 
\usepackage{multirow} 
\usepackage{booktabs} 
\usepackage[flushleft]{threeparttable} 
\usepackage{makecell}
\usepackage[section]{placeins}
\usepackage{newunicodechar}
\newunicodechar{−}{-}

\draftfalse
\journalname{JGR: Machine Learning and Computation}

\begin{document}

\title{A Machine Learning Framework for Predicting Microphysical Properties of Ice Crystals from Cloud Particle Imagery}

%
%

\authors{J. Ko\affil{1}, J. Y. Harrington\affil{2}, K. J. Sulia\affil{3}, V. Przybylo\affil{3}\thanks{Current affiliation: National Grid, Waltham, MA USA}, M. van Lier-Walqui\affil{4,5}, and K. D. Lamb\affil{6}}

\affiliation{1}{Columbia Climate School, Columbia University, New York, NY USA}
\affiliation{2}{Department of Meteorology \& Atmospheric Science, Pennsylvania State University, University Park, PA USA}
\affiliation{3}{Atmospheric Science Research Center, State University of New York (SUNY), Albany, NY USA}
\affiliation{4}{Goddard Institute for Space Studies, National Aeronautics \& Space Administration (NASA), New York, NY USA}
\affiliation{5}{Center for Climate Systems Research, Columbia University, New York, NY USA}
\affiliation{6}{Department of Earth \& Environmental Engineering, Columbia University, New York, NY USA}

\correspondingauthor{Joseph Ko}{jk4730@columbia.edu}


\begin{keypoints}
\item A machine learning pipeline to predict 3D microphysical properties from 2D imagery is developed using synthetic ice crystals.
\item Stereo view models were also trained and evaluated to quantify the marginal benefits of an additional view for predicting 3D properties.
\item This framework lays the foundation for predicting 3D microphysical properties of ice crystals from in situ optical imagery (e.g., CPI).
\end{keypoints}

%
%

\begin{abstract}
The microphysical properties of ice crystals are important because they significantly alter the radiative properties and spatiotemporal distributions of clouds, which in turn strongly affect Earth's climate. However, it is challenging to measure key properties of ice crystals, such as mass or morphological features. Here, we present a framework for predicting three-dimensional (3D) microphysical properties of ice crystals from in situ two-dimensional (2D) imagery. First, we computationally generate synthetic ice crystals using 3D modeling software along with geometric parameters estimated from the 2021 Ice Cryo-Encapsulation Balloon (ICEBall) field campaign. Then, we use synthetic crystals to train machine learning (ML) models to predict effective density ($\rho_{e}$), effective surface area ($A_e$), and number of bullets ($N_b$) from synthetic rosette imagery. When tested on unseen synthetic images, we find that our ML models can predict microphysical properties with high accuracy. For $\rho_{e}$ and $A_e$, respectively, our best-performing single view models achieved $R^2$ values of 0.99 and 0.98. For $N_b$, our best single view model achieved a balanced accuracy and F1 score of 0.91. We also quantify the marginal prediction improvements from incorporating a second view. A stereo view ResNet-18 model reduced RMSE by 40\% for both $\rho_e$ and $A_e$, relative to a single view ResNet-18 model. For $N_b$, we find that a stereo view ResNet-18 model improved the F1 score by 8\%. This work provides a novel ML-driven framework for estimating ice microphysical properties from in situ imagery, which will allow for downstream constraints on microphysical parameterizations, such as the mass-size relationship. 
\end{abstract}

\section*{Plain Language Summary}
The physical properties of ice crystals influence the overall behavior of clouds and subsequently their impacts on weather and climate. However, it is difficult to measure certain properties of ice crystals, such as mass, in real clouds. In this work, we develop a methodology to predict important properties of ice crystals based on 2D images of crystals taken in real clouds. Specifically, we train a computer model to predict the mass, surface area, and the number of bullets (i.e., ``spikes") of individual crystals, given crystal images captured with a research-grade instrument called a cloud particle imager. We find that our trained computer models are able to predict the properties of crystals with high skill. Predicting ice crystal properties accurately and efficiently will allow for the downstream improvement of cloud representations in computer models of the Earth's atmosphere.

%
%
\section{Introduction}

Clouds exert significant influence on Earth's energy balance and water cycle. However, it is difficult to represent clouds accurately within physics-based weather and climate models because of the complex microphysical properties and processes that govern cloud behavior \cite{morrisonConfrontingChallengeModeling2020}. Ice clouds present a particularly acute challenge because ice habit (i.e., shape) has non-trivial impacts on particle-radiation interactions, particle fall speeds, and microphysical process rates \cite{marshallTHEORYSNOWCRYSTALHABIT1954, atlasRadarRadiationProperties1995, baileyComprehensiveHabitDiagram2009, shimaPredictingMorphologyIce2020, chandrakarWhatControlsCrystal2024} and can vary independently of particle size and mass. These microphysical effects translate to macroscopic impacts on important bulk processes such as cloud-radiation interactions, cloud lifetime, and precipitation. 

Cirrus (ice) clouds are estimated to cover more than half of the Earth's surface at any given time, yet their radiative impacts are poorly constrained. For example, \citeA{jarvinenAdditionalGlobalClimate2018} found that ice crystal complexity alone can induce an additional shortwave cooling effect of -1.12 \unit{W.m^{-2}}, which is approximately 7\% of the estimated global mean shortwave cloud radiative effect. By conducting sensitivity studies using NCAR's Community Earth System Model (CESM), \citeA{wangImpactCloudIce2020} found that the habit-dependent effective size of ice and snow can account for changes in climate sensitivity (i.e., change in mean surface temperature in response to doubling carbon dioxide concentrations) between -6.2\% and 12.3\%. \citeA{yangEffectObservedVertical2012} used a cloud resolving model coupled to a radiative transfer scheme and found that habit representation can impact upwelling and down-welling radiative fluxes by about 12 and 16 \unit{W.m^{-2}}, respectively. \citeA{sullivanIceMicrophysicalProcesses2021} found that inconsistent treatment of ice crystal size between the microphysics and radiation schemes can alter cloud-radiative heating by a factor of four and mean infrared cooling by 30 \unit{W.m^{-2}} over the Asian Monsoon region. \citeA{wendischEffectsIceCrystal2007} performed radiative transfer modeling in combination with in situ measurements from the CRYSTAL-FACE field measurement campaign and estimated that ice habit can impact broadband thermal IR irradiance for high, optically thin cirrus clouds by up to 70\%. 

In addition to radiative uncertainties, ice habit representation can impact the prediction of precipitation. For example, \citeA{sterzingerEffectsIceHabit2021} found that variable-habit simulations of orographic snowfall produced on average 14\% more precipitation than fixed-habit simulations. In another study, \citeA{jensenImpactsIceParticle2018} show that a habit-evolving microphysics scheme (ISHMAEL) improved the prediction of ice water content and surface precipitation. From an experimental angle, \citeA{oraltayMeltingLayerLaboratory2005} conducted single-particle laboratory measurements of melting ice with various habits to highlight the wide range of melting behavior of ice crystals, which impacts downstream precipitation. 

\begin{figure}
    \centering
    \noindent\includegraphics[width=\textwidth]{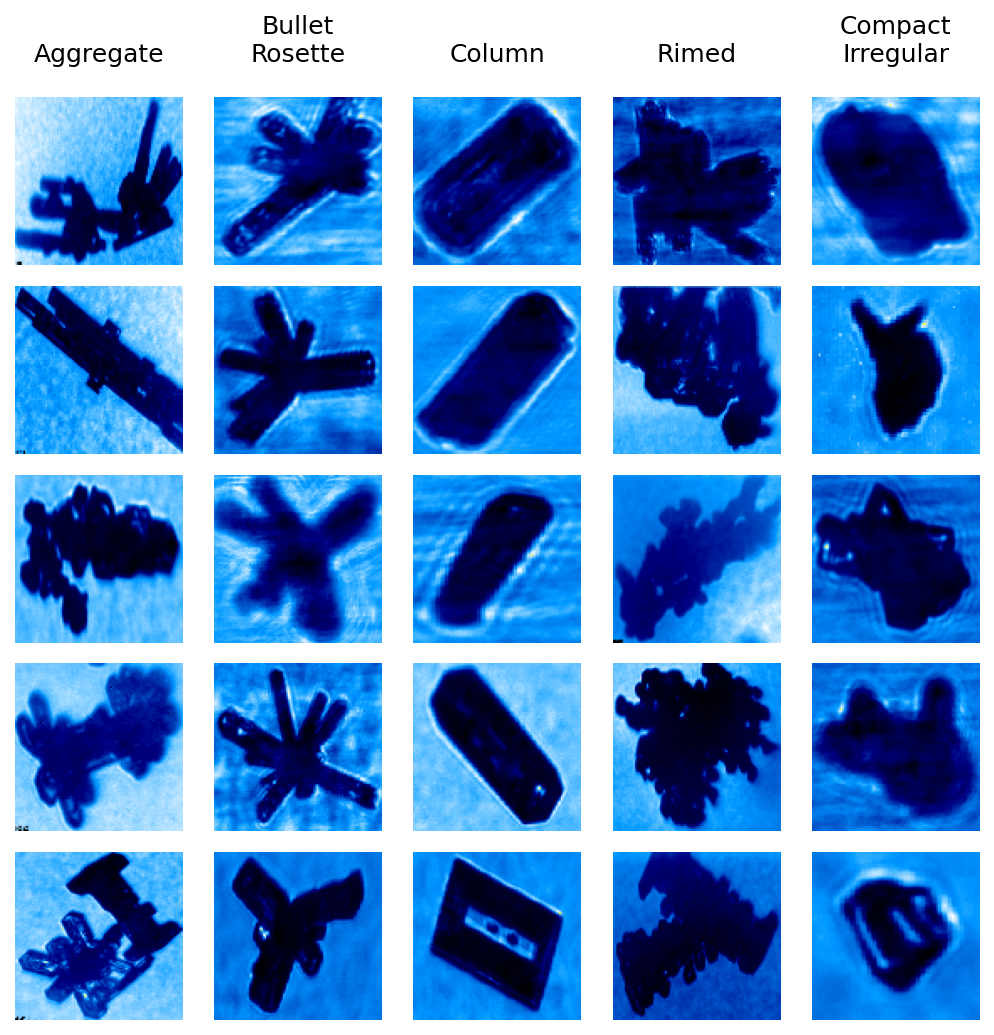}
    \caption{Examples of CPI imagery categorized by a different habit in each column. The CPI has a pixel resolution of \SI{2.3}{\micro\meter} and a size range of 15 to \SI{2500}{\micro\meter}, which allows for a wide variety of shapes and morphological details to be captured.}
    \label{fig:cpi}
\end{figure}

A first step towards decreasing uncertainty associated with ice habit is constraining habit-relevant properties using real-world observations. More specifically, in situ optical imaging of cloud particles from aircraft-based field campaigns in the past few decades allows for the estimation of cloud particle size distributions and habit compositions in diverse clouds types and atmospheric conditions. In situ optical imaging probes, such as the cloud particle imager (CPI), have been crucial to quantifying and constraining ice habit diversity, ice size distributions, and important dimensional relationships of ice, such as mass-size (m-D) and cross-sectional area-size (A-D) relationships \cite{brownImprovedMeasurementsIce1995, mcfarquharProcessingIceCloud2017, schmittDimensionalCharacteristicsIce2010, leroyIceCrystalSizes2016, lawsonSituObservationsMicrophysical2006, erfaniDevelopingBoundingIce2016, bakerImprovementDeterminationIce2006, jacksonDependenceIceMicrophysics2012}. However, the 2D nature of crystal images requires assumptions about the mapping between 2D length scales (e.g., maximum dimension length) and 3D related properties (e.g., mass, surface area, geometric features). For instance, \citeA{jiangWhatCanWe2017, jiangShapesFallOrientations2019} show that 2D projected aspect ratios and oblate spheroids are not generally representative of actual aggregate shapes. Examples of different ice crystals from the CPI are shown in Figure \ref{fig:cpi} for reference. 

In this work, we present a machine learning (ML) based framework to infer 3D attributes of ice crystals from 2D imagery. ML has been shown to be highly effective in computer vision tasks across diverse domains ranging from medical imaging \cite{ericksonMachineLearningMedical2017, gigerMachineLearningMedical2018} to remote sensing \cite{maxwellImplementationMachinelearningClassification2018, camps-vallsMachineLearningRemote2009}, and more specifically, the 2D-to-3D problem is a highly active field of research within the computer graphics and vision communities \cite{samavatiDeepLearningbased3D2023, hanImageBased3DObject2021}. Given the large volume of publicly available CPI data, and the strong precedence of computer vision for these types of tasks, ML is particularly well suited for this challenge. Using computationally generated ice crystal models (hereafter referred to as ``synthetic" crystals), we train ML models to predict the mass, surface area, and number of bullet arms of rosette crystals from CPI images in a supervised manner. 

We focus on rosette crystals in this study because rosettes constitute a substantial fraction of ice in cirrus clouds. For example, \citeA{lawsonSituObservationsMicrophysical2006} and \citeA{lawsonReviewIceParticle2019} found that rosette-like particles constituted over 50\% of the surface area and mass of ice particles that were larger than 50 $\mu$m in cirrus clouds. Rosettes also have complex radiative properties due to their shape and their physical properties that are not yet properly represented in models of pristine ice formation \cite{pokrifkaEffectiveDensityDerived2023}. Although this work is focused on rosettes, our framework can be extended to other ice habits in future studies. 

This manuscript is organized into four subsequent sections. Section 2 describes the data and methods used in this study. Section 3 presents the results from the different predictive models developed in this study. Section 4 discusses how these models can be used to inform and constrain ice microphysical parameterizations. Section 5 summarizes the main results and also discusses limitations and next steps.

\section{Data and Methods} \label{section_data_methods}

\subsection{Data description: synthetic ice crystals}
It is difficult to infer the 3D properties of in-cloud ice crystals from in situ measurements, especially with only a single 2D view. Multi-view instruments such as the multi-angle snowflake camera (MASC), 3-view cloud particle imager (3V-CPI), and particle habit imaging and polar scattering probe (PHIPS) can help resolve ice crystal morphology in more detail with additional simultaneous views \cite{garrettFallSpeedMeasurement2012, abdelmonemPHIPSHALOAirborne2016a, lawsonOverviewMicrophysicalProperties2001, lawson2DSStereoProbe2006}. However, the MASC is primarily designed to measure precipitating ice at ground-level, and there is less data available from 3V-CPI or PHIPS measurements relative to CPI measurements. Even with additional views, algorithms are ultimately needed to infer 3D information from 2D views. 

Tomographic imaging techniques, such as X-ray micro-computed tomography (micro-CT) and electron tomography, have been applied in various fields to reconstruct microscopic 3D structures, with resolutions ranging from $\sim$\unit{\nm} to $\sim$\unit{\um} \cite{ritmanMicroComputedTomographyCurrent2004, weylandElectronTomography2004}. For example, micro-CT has been used in various studies to characterize the 3D structure of individual snow particles \cite{ishimotoSnowParticlesExtracted2018, haffarXrayTomography3D2021}. In biological and material sciences, variants of electron tomography have been used to reconstruct the 3D structures of proteins, viruses, nanoparticles, and more \cite{midgleyElectronTomographyHolography2009, scottElectronTomography24angstrom2012, albrechtFastElectronTomography2020, yaoMolecularArchitectureSARSCoV22020}. However, in situ tomographic imaging of ice crystals in the atmosphere is currently not possible, mainly due to sampling restrictions and the relatively laborious process of developing high quality tomographic reconstructions. 

Even if we had the technology to explicitly resolve the 3D microstructure of in situ ice crystals, the sheer volume of tomographic data to process would be unwieldy to handle and it is not clear whether the extreme levels of morphological detail would be immediately helpful in the context of constraining microphysical parameterizations \cite{lambPerspectivesSystematicCloud2025}. In other words, the level of morphological detail required is entirely context-dependent, and for microphysical  modeling, statistical descriptions of microphysical properties and relationships at the population level are more pertinent than extremely precise descriptions of a limited number of crystals. 

\begin{figure}
    \centering
    \noindent\includegraphics[width=\textwidth]{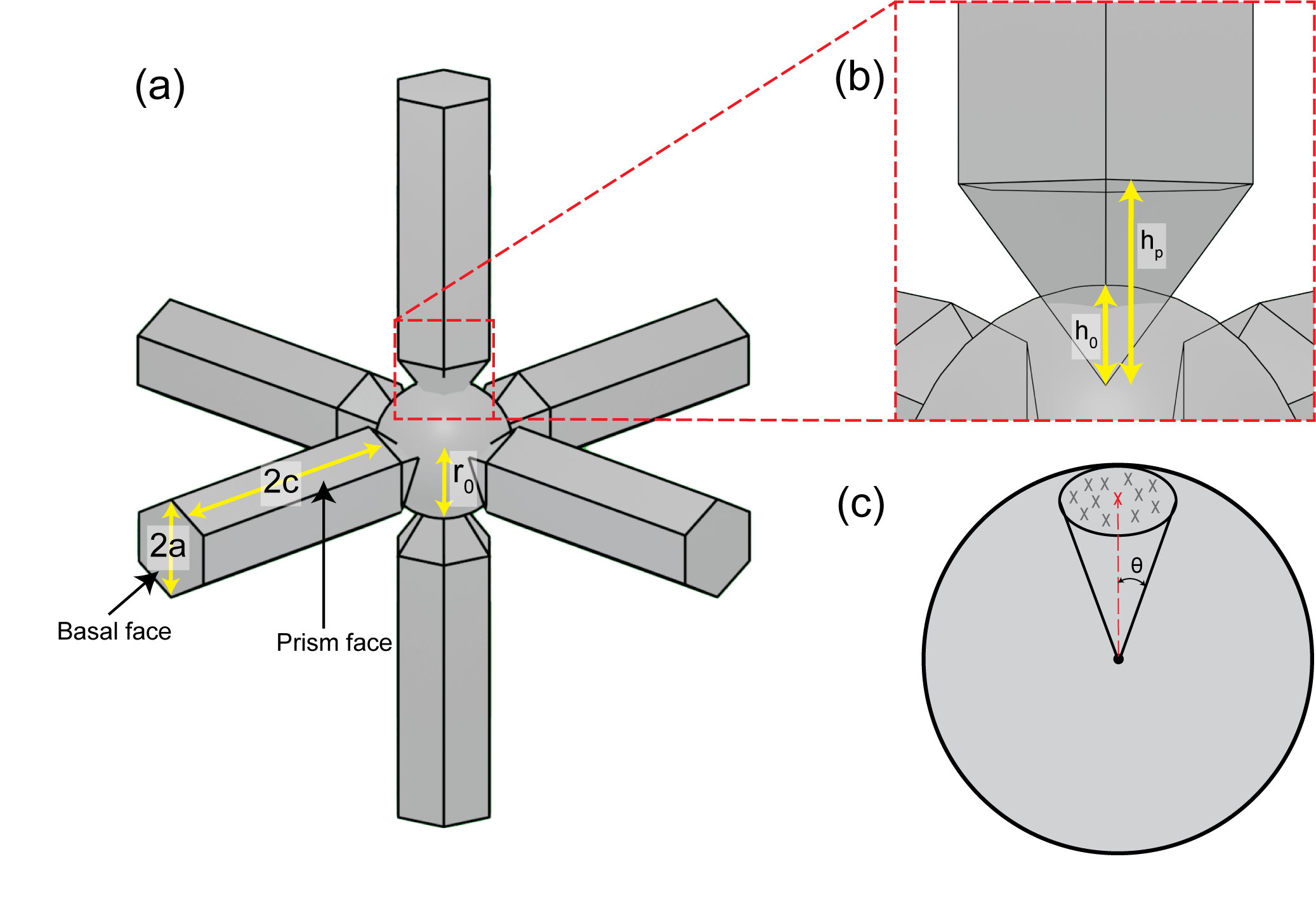}
    \caption{Illustration of synthetic rosette geometric parameters. (a) $a$ is the half-length of the bullet arm basal face, $c$ is the half-length of the bullet arm prism face, and $r_0$ is the radius of the center sphere. (b) $h_p$ is the height of the pyramidal tip of a bullet arm and $h_0$ is the depth of penetration of the tip into the center sphere. (c) $\theta$ describes the angle between the center and edge of a solid angle that defines the spherical cap on which the bullet arm is randomly placed. The red ``x" mark signifies the initial point on the sphere dictated by optimal spherical code. The black ``x" marks are illustrative examples of random samples on the spherical cap.}
    \label{fig:rosette}
\end{figure}

These limitations and challenges in obtaining 3D crystal information gives us the impetus to develop a scalable methodology that can efficiently infer 3D crystal properties from readily available CPI datasets spanning numerous NASA/DOE/NSF-funded airborne field campaigns conducted across the past few decades \cite{lawsonOverviewMicrophysicalProperties2001, houzeOlympicMountainsExperiment2017, dasIceCrystalHabit2025, mcfarquharSmallCloudParticle2013, jensenNASAAirborneTropical2017, lawsonMicrophysicsIcePrecipitation2015, przybyloClassificationCloudParticle2022}. Since we do not have corresponding 3D ground truth for CPI images, we computationally generate synthetic analogs of rosette crystals, based on \textit{a priori} geometric assumptions and stochastic perturbations of geometric parameters. The geometric assumptions presented in \citeA{pokrifkaEffectiveDensityDerived2023} were adapted to create a dataset of 3D rosette crystal models. The geometric parameters in our rosette model include: radius of the central sphere ($r_0$), half-length of the bullet basal face ($a$), bullet aspect ratio ($a/c$), height of the bullet pyramid tip ($h_p$), depth of bullet penetration into the central sphere ($h_0$), and number of bullets ($N_b$). Figure \ref{fig:rosette} illustrates these rosette geometric parameters in a schematic. 

\begin{figure}
    \centering
    \noindent\includegraphics[width=\textwidth]{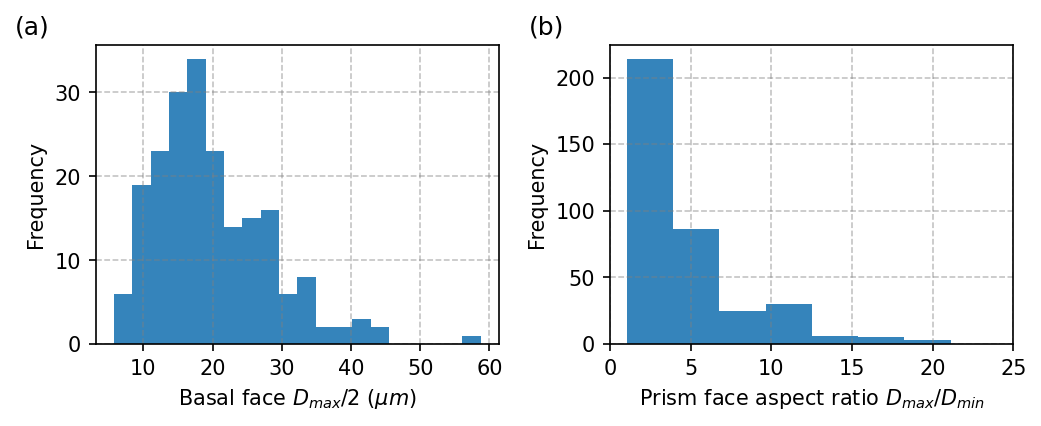}
    \caption{Distributions of (a) half-length of the basal face and (b) prism face aspect ratio from ICEBall SEM imagery.}
    \label{fig:iceball-dist}
\end{figure}

The ranges of geometric parameters were constrained using observations from the Ice Cryo-Encapsulation Balloon (ICEBall) field campaign conducted between October 16 and November 6, 2021, at the U.S. Department of Energy Atmospheric Radiation Measurement Southern Great Plains site \cite{harringtonIceCryoEncapsulationBalloon2023}. Measurements of ice crystal dimensions were taken using scanning electron microscopy (SEM) images of cryogenically preserved samples from the ICEBall campaign. Specifically, measurements of basal face length ($D_{max}$), prism face aspect ratio ($D_{max}$/$D_{min}$), and the number of bullet arms from samples taken between October 23 and 25 were used as constraints for our synthetic dataset. These dates were selected due to the prevalence of rosettes during this sampling period. The distributions of $D_{max}/2$ and $D_{max}$/$D_{min}$ from measured ICEBall imagery are shown in Figure \ref{fig:iceball-dist}. The range of $N_b$ was set to [4, 10] based on the minimum and maximum number of bullet arms observed in the ICEBall data. This $N_b$ range is also comparable to the range identified using PHIPS measurements in \citeA{wagnerLightScatteringMicrophysical2024}.

The initial distributions of geometric parameters for the synthetic crystals were controlled by five parameters: $a$, $c/a$, $f_{r_0}$, $f_{h_p}$, and $f_{h_0}$ (see Table \ref{table:params} for descriptions). The 5th and 95th percentile values of $D_{max}/2$ and $D_{max}$/$D_{min}$ were used to define the ranges for $a$ and $c/a$, respectively. The variables $f_{r_0}$, $f_{h_p}$, and $f_{h_0}$ refer to stochastic scaling factors for $r_0$, $h_p$, and $h_0$, and the ranges for these parameters were set to \SI{\pm 20}{\percent}. The variables $r_0$, $h_p$, and $h_0$ were diagnosed as a function of $N_b$ and $a$, as described in Equations \ref{eqn:r0} to \ref{eqn:h0}. In brief, $r_0$ is linearly parameterized as a function of $N_b$ and bound by $0.5 \cdot a$ and $a$; $h_p$ is linearly parameterized as a function of $N_b$ and bound by $r_0$ and $1.5 \cdot r_0$; $h_0$ is prescribed as half of $r_0$; and all three parameters are multiplied by their respective stochastic scaling factors. 

\begin{equation}
    \begin{gathered}
        r_0 = f_{r_0} \cdot (\beta_{r_0, 1} \cdot N_b + \beta_{r_0, 0}), \\
        \text{where}\ \beta_{r_0, 1} = \frac{(r_{0, max} - r_{0, min})}{(N_{b, max} - N_{b, min})} \\
        \beta_{r_0, 0} = r_{0, min} - \beta_{r_0, 1} \cdot N_{b, min} \\
        N_{b, min} = 4,\ N_{b, max} = 10,\ r_{0, min} = a/2, \ r_{0, max} = a
    \end{gathered}
    \label{eqn:r0}
\end{equation}
\hrule
\begin{equation}
    \begin{gathered}
        h_p = f_{h_p} \cdot (\beta_{h_p, 1} \cdot N_b + \beta_{h_p, 0}), \\
        \text{where}\ \beta_{h_p, 1} = \frac{(h_{p, max} - h_{p, min})}{(N_{b, max} - N_{b, min})}\ \\
        \beta_{h_p, 0} = h_{p, min} - \beta_{h_p, 1} \cdot N_{b, min} \\
        N_{b, min} = 4,\ N_{b, max} = 10,\ h_{p, min} = r_0, \ h_{p, max} = 1.5 \cdot r_0
    \end{gathered}
    \label{eqn:hp}
\end{equation}
\hrule
\begin{equation}
    h_0 = f_{h_0} \cdot \frac{r_0}{2}
    \label{eqn:h0}
\end{equation}

Latin hypercube sampling was then used to sample 200 combinations of \{$a$, $c/a$, $f_{r_0}$, $f_{h_p}$, $f_{h_0}$\}. The centered discrepancy metric was used to check that 200 samples would sufficiently cover the parameter space \cite{zhouMixtureDiscrepancyQuasirandom2013, virtanenSciPy10Fundamental2020}. This set of 200 parameter combinations was then repeated seven times, once for each discrete class of $N_b$ (4 to 10 bullet arms). This results in 1,400 combinations of \{$a$, $c/a$, $f_{r_0}$, $f_{h_p}$, $f_{h_0}$, $N_b$\}, which we refer to as ``base" geometric parameters. The base geometric parameter ranges used to generate our crystals are listed in Table \ref{table:params}.

\begin{table}[hbt!]
    \caption{Range of base geometric parameter values for synthetic rosettes.}
    \centering
    \begin{tabular}{l l l}
    \hline
     Parameter  & Description & Range of Values  \\
    \hline
      $r_0$  & Radius of center sphere &[a, b] \unit{\um} \\
      $a$  & Half-length of basal face & [9.44, 33.90] \unit{\um} \\
      $c/a$  & Aspect ratio of bullet arm & [1.19, 5.75]  \\
      $N_b$  & Number of bullet arms & [4, 10]   \\
      $f_{r_0}$ & Stochastic scaling factor for $r_0$ & [0.8, 1.2]\\
      $f_{h_p}$ & Stochastic scaling factor for $h_p$ & [0.8, 1.2] \\
      $f_{h_0}$ & Stochastic scaling factor for $h_0$ & [0.8, 1.2] \\
    \hline
    \end{tabular}
    \label{table:params}
\end{table}

After generating 1,400 combinations of base geometric parameters, 50 random variants were created from each base state, resulting in 70,000 randomly perturbed parameter sets. These random variations consisted of three different perturbations: scaling of the basal face, scaling of the prism face, and placement of the bullet arms. The bullets locations are initially distributed on the center sphere using spherical codes specified in \citeA{sloaneSphericalCodes2000}, which maximizes the minimum distance between points on a sphere. After initial placement of the bullets on the sphere according to the procedures described above, each bullet's aspect ratio and placement on the sphere are randomly perturbed within a prescribed range. 

The scaling of the basal and prism faces was controlled by scaling factors $f_a$ and $f_c$, respectively. The limits of $f_a$ and $f_c$ were set to [0.8, 1.2], which means that the basal and prism face lengths for each bullet arm were randomly scaled within \SI{\pm 20}{\percent}. The perturbations of bullet placement on the center sphere were controlled by the parameter $\theta$, where $\theta$ is the angle between the center and edge of a solid angle that defines a spherical cap on which the bullet arm is randomly placed (see Figure \ref{fig:rosette}c). $\theta$ was prescribed as a function of $N_b$ described by Equation \ref{eqn:theta}, where $\phi_{N_b}$ is the minimal angle of separation between two points on a sphere given by spherical code \cite{sloaneSphericalCodes2000}, corresponding to $N_b$ = [4, 10] in monotonically increasing order. Bullets were placed on the sphere such that the prism axis (i.e., c-axis) is normal to the sphere surface.

\begin{equation}
    \begin{gathered}
        \theta = \frac{\phi_{N_b}}{3}, \\
        \text{where}\ \phi_{N_b} = \{109.47,\, 90.00,\, 90.00,\, 77.87,\, 74.86,\, 70.53,\, 66.15\}
        \label{eqn:theta}
    \end{gathered}
\end{equation}

\begin{figure}
    \centering
    \noindent\includegraphics[width=\textwidth]{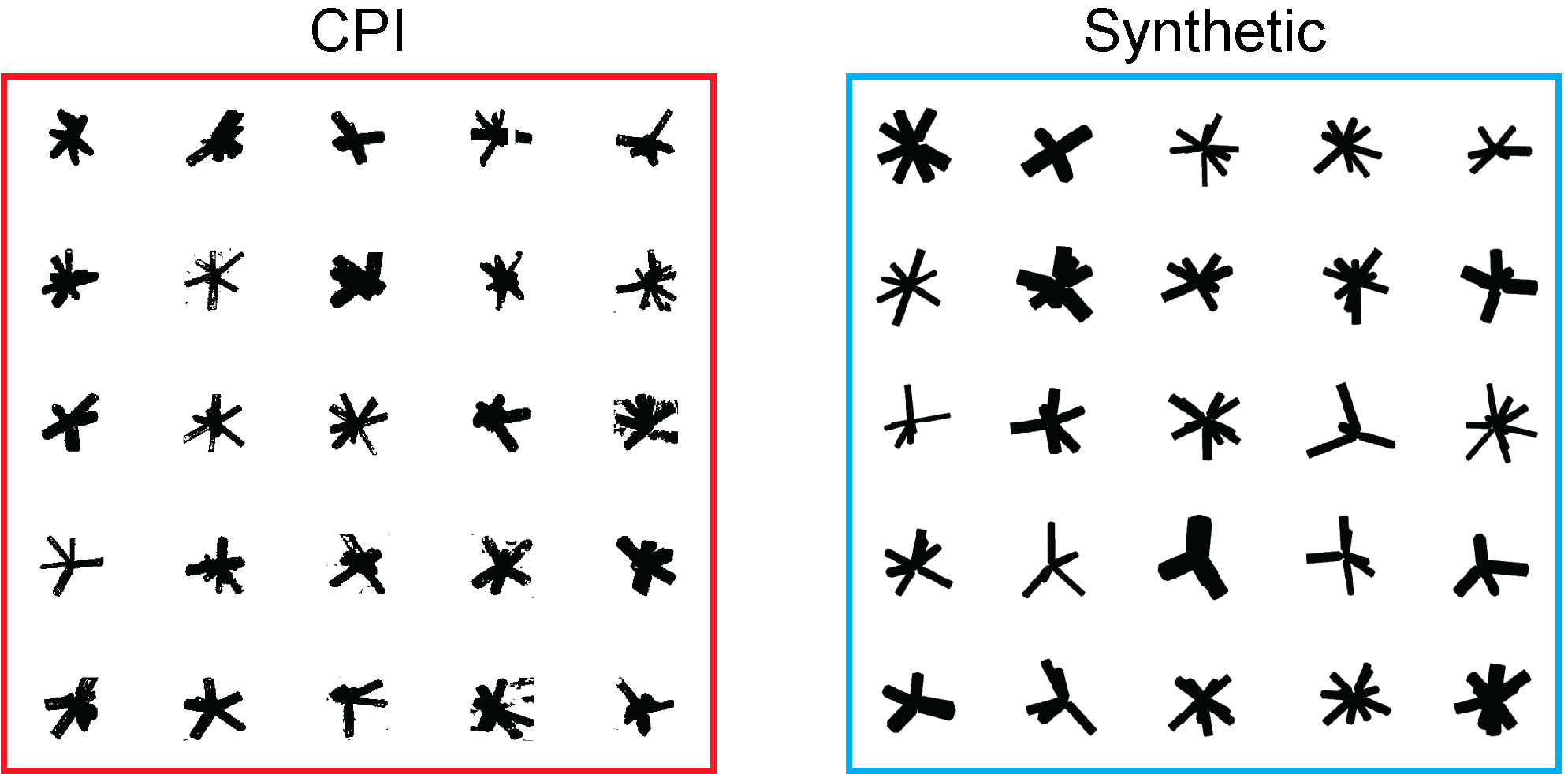}
    \caption{Representative samples of masked projections from CPI images (left) and synthetic crystals (right).}
    \label{fig:cpi-v-synth}
\end{figure}

In total, this data generation process resulted in a 3D dataset of 70,000 randomly perturbed rosette crystal models. For each crystal, the attributes listed in Table \ref{table:crystal-attributes} were calculated and 100 random 2D projections were rendered, resulting in a 2D image dataset of 7 million samples. We used orthographic projections to create single-channel, binary masked images, each with a resolution of 224x224 pixels. Sample renders from our 2D dataset are shown in Figure \ref{fig:cpi-v-synth}. Qualitatively, we find our synthetic rosette projections to be visually similar to actual CPI projections. A more realistic approach would be to replicate the particle-light interactions to emulate the charge-coupled device sensor in the optical imaging probes. However, this method is beyond the scope of the current study but is recommended for potential future work. Additionally, we found that even binary masks were sufficient to predict 3D attributes with high skill, suggesting that the outline of particle shape is sufficient to guide ML models to effectively predict 3D attributes of interest. 

10\% of the 7 million images, or 700,000 samples, were used for the training and evaluation of models presented in this study. This decision was made in order to facilitate efficient model training, and we also observed a diminishing return on increasing sample size for model performance. The final ML-ready dataset is an image dataset consisting of 700,000 2D projections, along with an accompanying tabular dataset containing the corresponding geometric parameters, unique particle ID, and calculated input features for each sample. Eight geometrically-relevant, input features are calculated for each image to be used as input for traditional ML models that require feature-engineered inputs. These features are described further in Section \ref{section:task1}.

\begin{table}[hbt!]
    \caption{Descriptions of calculated crystal attributes}
    \centering
    \begin{tabular}{l l p{8cm}}
    \hline
     Attribute  & Units & Description  \\
    \hline
        $V$ & \unit{m^3} & Volume of crystal \\
        $A$ & \unit{m^2} & Surface area of crystal \\
        $V_{mbs}$ & \unit{m^3} & Volume of minimal bounding sphere \\
        $A_{mbs}$ & \unit{m^2} & Surface area of minimal bounding sphere \\
        $\rho_{e}$ & unitless & Effective density defined as volume of crystal divided by volume of minimal bounding sphere i.e., $V / V_{mbs}$\\
        $A_{e}$ & unitless & Effective surface area defined as surface area of crystal divided by surface area of minimal bounding sphere i.e., $A/ A_{mbs}$\\
    \hline
    \end{tabular}
    \label{table:crystal-attributes}
\end{table}

In addition to the 2D single view dataset, two different 2D stereo view datasets were created, specifically tailored to train models that can make predictions based on stereo optical imagery (see Section \ref{section:task2} for more details). Each of the two 2D stereo view datasets were constructed in a similar fashion to the 2D single view dataset, except each data record consists of a pair of stereo imagery. Each 2D stereo view dataset consists of stereo image pairs with specific viewing angles of \ang{90} and \ang{120} to match the viewing angle of two different stereo optical imagers, the 2D-S \cite{lawson2DSStereoProbe2006} and the PHIPS \cite{abdelmonemPHIPSHALOAirborne2016a}, respectively. The first image of each image pair in both 2D-S and PHIPS stereo datasets are identical and index-matched to the single view 2D dataset for consistency. 

3D models were generated using CadQuery, an open-source Python package for parametric CAD modeling, built on top of OpenCascade \cite{auCadQueryCadqueryCadQuery2024}. Generated models were saved as mesh files in STL format, and projections were subsequently taken using PyVista, an open-source 3D modeling package \cite{sullivanPyVista3DPlotting2019, schroederVisualizationToolkit4th2006}.

\subsection{Learning tasks and model descriptions}
\begin{figure}
    \centering
    \noindent\includegraphics[width=\textwidth]{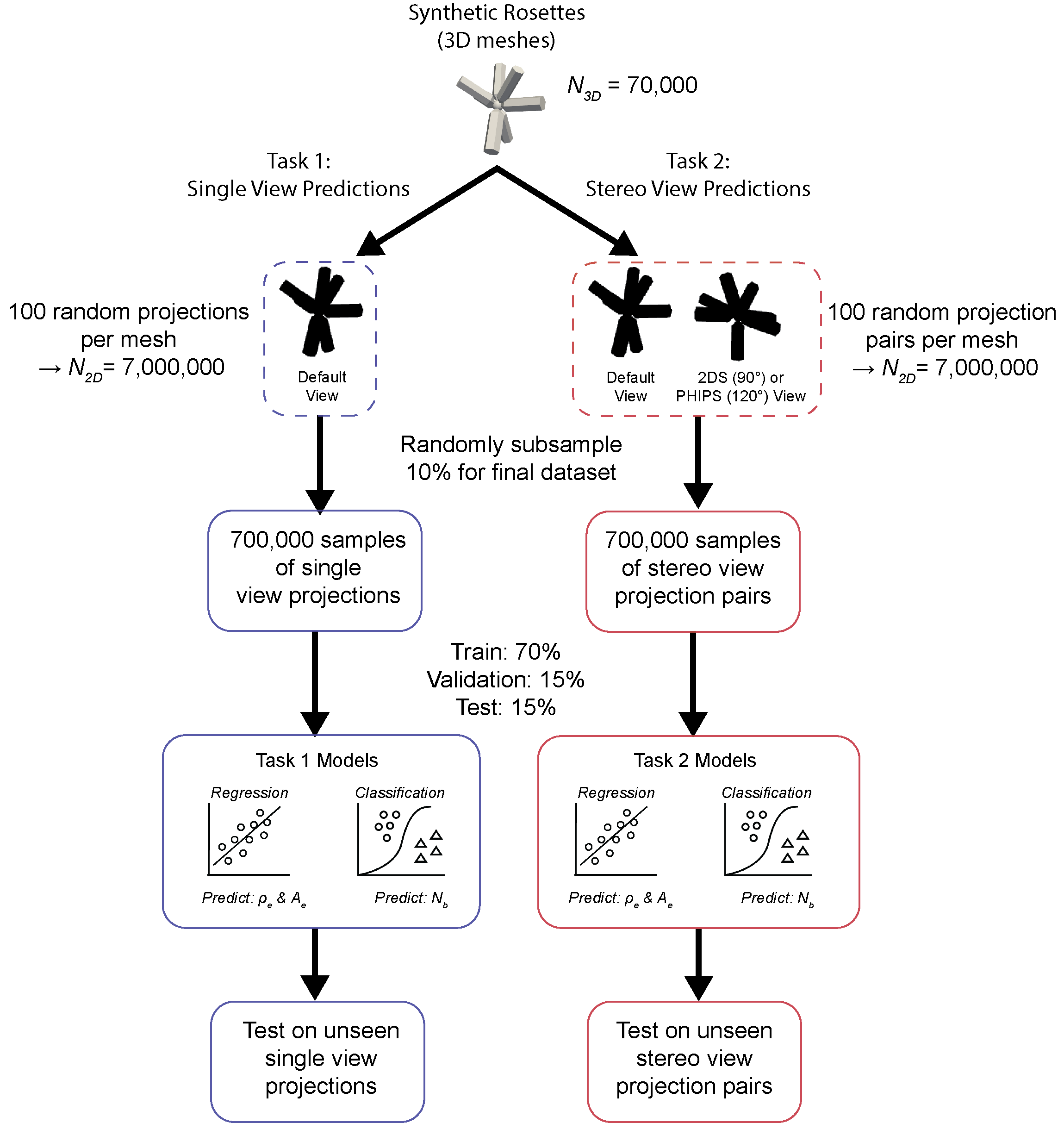}
    \caption{ML pipeline for Tasks 1 and 2.}
    \label{fig:pipeline}
\end{figure}

The machine learning tasks performed with this data and the model pipeline details are described in the subsequent sections (see Figure \ref{fig:pipeline} for visual overview). With our 2D datasets, we train supervised machine learning models to perform two main tasks, which are designed in the context of common interpretation tasks for in situ imaging data from various airborne field campaigns. The first task is to predict $\rho_{e}$, $A_e$, and $N_b$, given a single 2D image of a rosette. The second task is to predict $\rho_{e}$, $A_e$, and $N_b$, given a stereo pair of 2D images of the same rosette. A training, validation, and test split of 70\%, 15\%, and 15\% was used on the final 700,000 sample dataset, and these datasets were kept consistent between all models.

\subsubsection{Task 1: prediction from single images} \label{section:task1}
Task 1 can be further sub-divided into regression and classification tasks. Regression models are trained to predict continuous values of $\rho_{e}$ and $A_{e}$, while classification models are trained to predict discrete values of $N_b$.

For our regression subtask, we train our models to predict $\rho_{e}$ and $A_{e}$, which makes our models invariant to crystal size. An important caveat to emphasize is that all input images were re-sized to 224x224 pixels. Future work may investigate the use of models that are agnostic to input image resolution, but that is outside the scope of this work. Once $\rho_{e}$ and $A_{e}$ are predicted by our models, $m$ and $A$ can be calculated using the 2D maximum dimension of the crystal as a characteristic length scale of the enclosing sphere. 

Five different models were trained and evaluated for the single view regression subtask: linear regression (as a baseline), random forest regression, multi-layer perceptron (MLP),``vanilla" convolutional neural network (CNN), and ResNet-18 \cite{heDeepResidualLearning2015}. For the single view classification subtask, the equivalent classifier algorithms were used, except logistic regression was used as our baseline linear classifier model. For the linear regression, logistic regression, random forest, and MLP models, eight features were used as input features: aspect ratio, elliptical aspect ratio, number of extreme points, contour area, area ratio, complexity, and circularity. These features are a subset of common particle geometric characteristics specified in Section 4b of \citeA{przybyloClassificationCloudParticle2022}, and further detailed in \ref{App:A}. For CNN and ResNet-18 models, no feature engineering was required since the model architectures are inherently designed for image recognition tasks and therefore directly ingest images as inputs. Further details regarding neural network architectures are described in \ref{App:B}.

We chose to limit the scope of candidate ML models in this study because our goal was to establish a general framework, not necessarily to maximize model performance. Therefore, we chose five canonical supervised learning algorithms with varying levels of complexity to demonstrate the feasibility of our general ML pipeline. Starting from the most classical and simple model, we chose linear regression and logistic regression as linear baseline models. Increasing in model complexity, we chose the random forest algorithm because of its popularity and widespread use since its inception more than two decades ago \cite{breimanRandomForests2001}. In short, random forest is an ensemble, tree-based learning algorithm that can be used for both classification and regression, and it has been shown to be highly effective and efficient for predictive tasks using tabular data \cite{grinsztajnWhyTreebasedModels2022, louppeUnderstandingRandomForests2015}. 

Finally, we also used three different deep learning algorithms: MLP, CNN, and ResNet-18. The MLP serves as a basic deep learning baseline. The CNN incorporates convolutional filters in its architecture to improve its performance on image-based tasks. ResNet further improves upon the vanilla CNN architecture by incorporating residual blocks and skip connections, which help avoid the pitfalls of degradation (i.e., exploding or vanishing gradients) during training of deeper neural networks. We refer readers to the following references for further details regarding neural networks and deep learning for image recognition: \citeA{chengNeuralNetworksReview1994, lecunDeepLearning2015, guoDeepLearningVisual2016, lecunGradientbasedLearningApplied1998, heDeepResidualLearning2015}.

A visual schematic of the general ML pipeline is shown in Figure \ref{fig:pipeline}. Regression models were trained to predict $\rho_e$ and $A_e$ simultaneously, and classification models were trained to predict $N_b$. Scikit-learn \cite{pedregosaScikitlearnMachineLearning2011} was used to train and evaluate the linear regression, logistic regression, and random forest models. PyTorch \cite{paszkeAutomaticDifferentiationPyTorch2017} was used to train and evaluate the MLP, CNN and ResNet models. 

\subsubsection{Task 2: prediction from stereo image pairs} \label{section:task2}
Task 2 was conducted in a similar fashion to Task 1, except two images of the same crystal were used as inputs instead of a single image. The main purpose of Task 2 was to quantify the advantage of an additional view for the prediction of $\rho_e$, $A_e$, and $N_b$. The real-world relevance of this task is in the context of stereo view optical imagers, such as the 2D-S probe \cite{lawson2DSStereoProbe2006} and the PHIPS \cite{abdelmonemPHIPSHALOAirborne2016a, schnaiterPHIPSHALOAirborneParticle2018}. These instruments have respective viewing angles of \ang{90} and \ang{120}. Random stereo projections of synthetic crystals were taken at \ang{90} and \ang{120}, to emulate the viewing angle of these instruments. For consistency, each pair uses the same first view projection contained in the original single view dataset. The same training and test split ratios from Task 1 was used for Task 2. 

For Task 2, only the linear baseline and ResNet-18 models were trained. The least and most complex models were selected in order to demonstrate the range of marginal benefits from adding a second view. Although stereo view versions of the random forest, MLP, and CNN models could also be trained, we found it unnecessary for the sake of the analysis presented here in this study. For the linear regression, the relevant features (aspect ratio, elliptical aspect ratio, etc.) were extracted for both stereo images and then concatenated into a single input feature array, resulting in a 16-dimensional input array. For ResNet-18, the stereo input images were concatenated along the channel dimension to create a stacked, 2-channel image input. The right branch of Figure \ref{fig:pipeline} illustrates the pipeline for Task 2.

\section{Results}
All results presented here are based on predictions made on the synthetic test dataset, which was completely isolated from the training process. Although the ultimate goal is to use this predictive framework with CPI data, we do not present evaluation statistics on CPI images here because there is no corresponding ground truth readily available for model evaluation. Furthermore, we recognize that many fine-scale morphological details, such as hollowing and surface irregularities, are not represented in our idealized models. These limitations and plans for future improvements are discussed in Section \ref{conclusion}. Results and implications of applying these ML models on real CPI data will be detailed in a follow-up study, but this is outside the scope of the present study, which focuses on developing a general methodology. The results from regression models predicting $\rho_e$ and $A_e$ are presented first. Then, the results from the classification models predicting $N_b$ are described. In each respective section below, the results from single view models (Task 1) are presented first, followed by results from the stereo view models (Task 2).

\subsection{Predicting effective density ($\rho_e$) and effective surface area ($A_e$)}
\begin{figure}
    \centering
    \noindent\includegraphics[width=\textwidth]{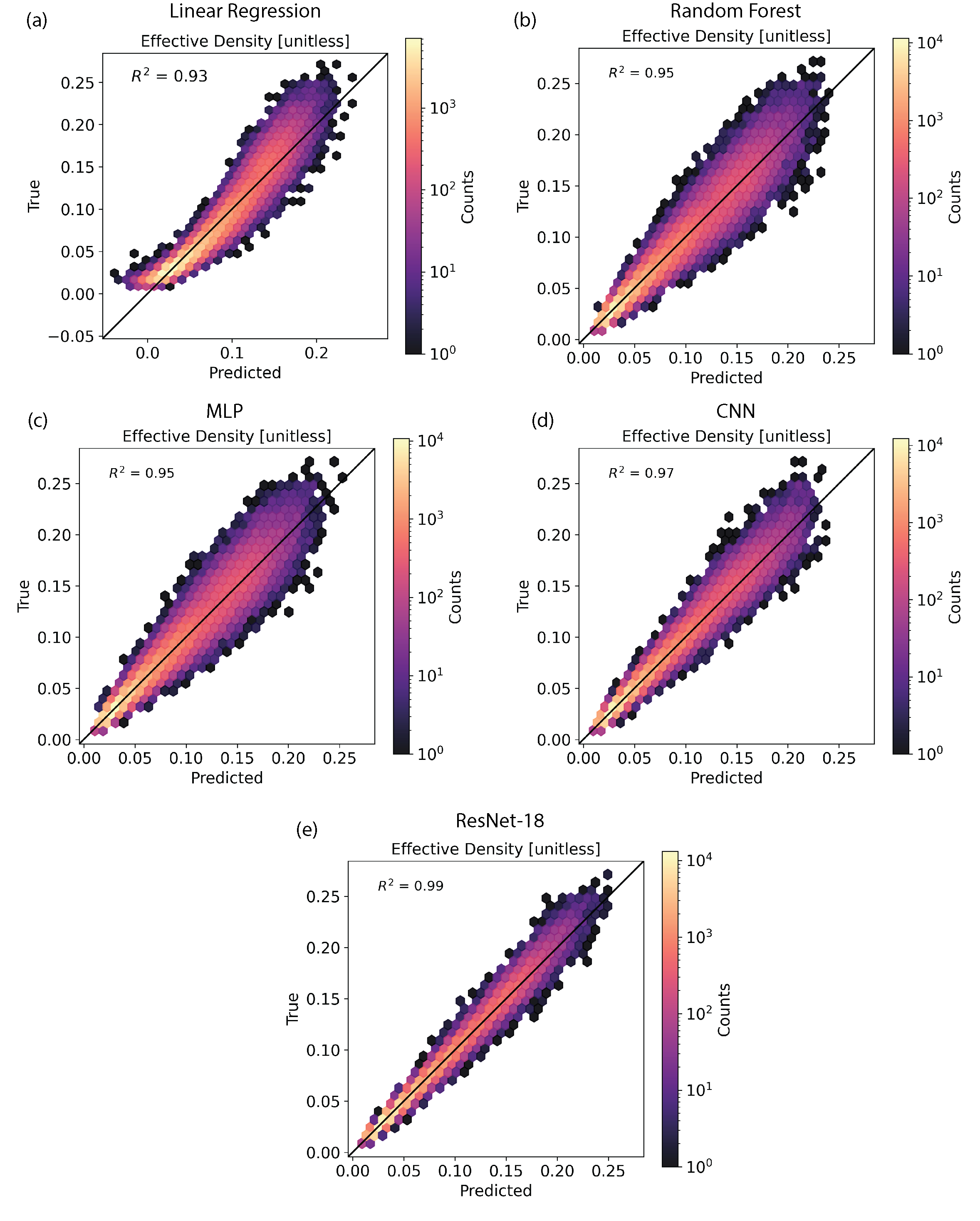}
    \caption{True vs. predicted values for $\rho_e$ for each of the five different single view regression models.}
    \label{fig:scatter-single-view-rho}
\end{figure}

\begin{figure}
    \centering
    \noindent\includegraphics[width=\textwidth]{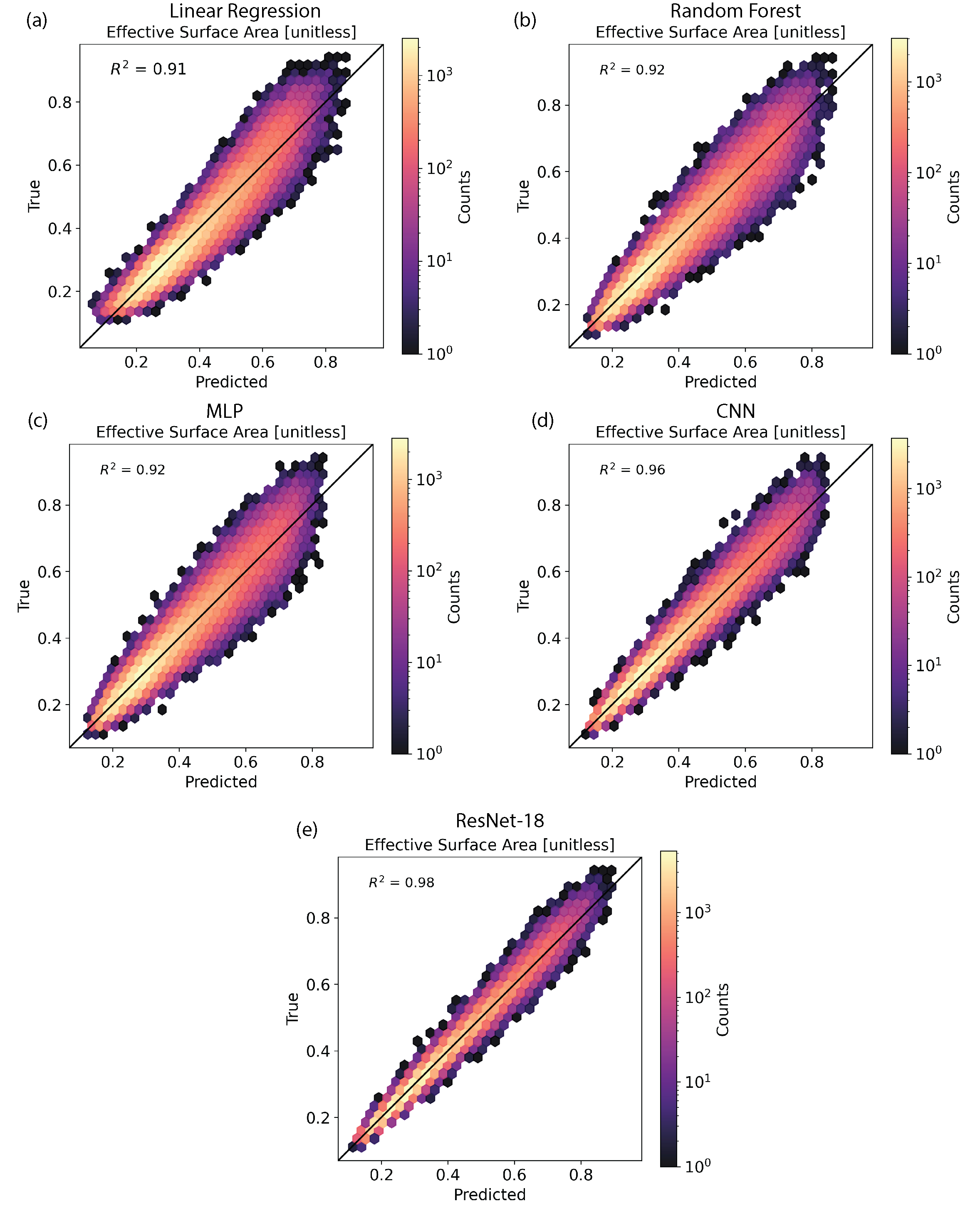}
    \caption{True vs. predicted values for $A_e$ for each of the five different single view regression models.}
    \label{fig:scatter-single-view-sa}
\end{figure}

Table \ref{table:regression_metrics} summarizes the performance of all regression models on the test dataset and Figures \ref{fig:scatter-single-view-rho} and \ref{fig:scatter-single-view-sa} show the scatter between actual and predicted values for all Task 1 (single view) regression models. For Task 1, ResNet-18 performed the best for the prediction of both $\rho_{e}$ and $A_{e}$, with $R^2$ values of 0.99 and 0.98 for $\rho_{e}$ and $A_{e}$. In comparison, the linear regression model resulted in $R^2$ values of 0.93 and 0.91, respectively, for $\rho_{e}$ and $A_{e}$. The random forest, MLP, and CNN models respectively resulted in $R^2$ values of 0.95, 0.95, and 0.97 for $\rho_{e}$; and 0.92, 0.96, and 0.98 for $A_{e}$. The root mean square error (RMSE) and mean absolute error (MAE) were also calculated for each model as additional evaluation metrics. The RMSE measures the mean difference between predicted and actual values, and the MAE measures the mean magnitude of errors. Formulas for RMSE and MAE can be found in \ref{App:C}. The RMSE for $\rho_{e}$ ranged between \num{4.75E-03} and \num{1.17E-02}, with ResNet-18 and linear regression having the lowest and highest RMSE, respectively. Similarly, the RMSE for $A_{e}$ ranged between \num{1.86E-02} and \num{4.45E-02}. MAE ranged between \num{5.48E-02} and \num{9.19E-02} for $\rho_e$, and between \num{1.15E-01} and \num{1.85E-01} for $A_e$, with ResNet-18 and linear regression having the lowest and highest MAE values. 

In general, all regression models demonstrated moderate to high skill in predicting both $\rho_e$ and $A_e$ for Task 1. Even the linear regression baseline model is able to predict $\rho_e$ and $A_e$ with a modest level of accuracy, given its simplicity. However, ML-based regression models clearly outperformed the linear baseline, which was expected given the ability of the ML models to capture more complex, non-linear relationships between inputs and outputs. This is particularly apparent in the lower and upper ranges of $\rho_e$ values (see Figure \ref{fig:scatter-single-view-rho}a), where there is a clear positive bias for the linear regression models. This positive bias is also apparent for $A_e$ (see Figure \ref{fig:scatter-single-view-sa}a), but to a lesser extent. In other words, the scatter is not symmetric about the 1:1 line, indicating a bias. The other ML models are able to largely mitigate this bias. The relative improvements from the linear baseline to more complex ML models is also clear when visually inspecting the spread about the 1:1 line in Figures \ref{fig:scatter-single-view-rho} and \ref{fig:scatter-single-view-sa}, with the spread gradually decreasing from subplots (a) to (e), for both $\rho_e$ and $A_e$.

\begin{figure}
    \centering
    \noindent\includegraphics[width=\textwidth]{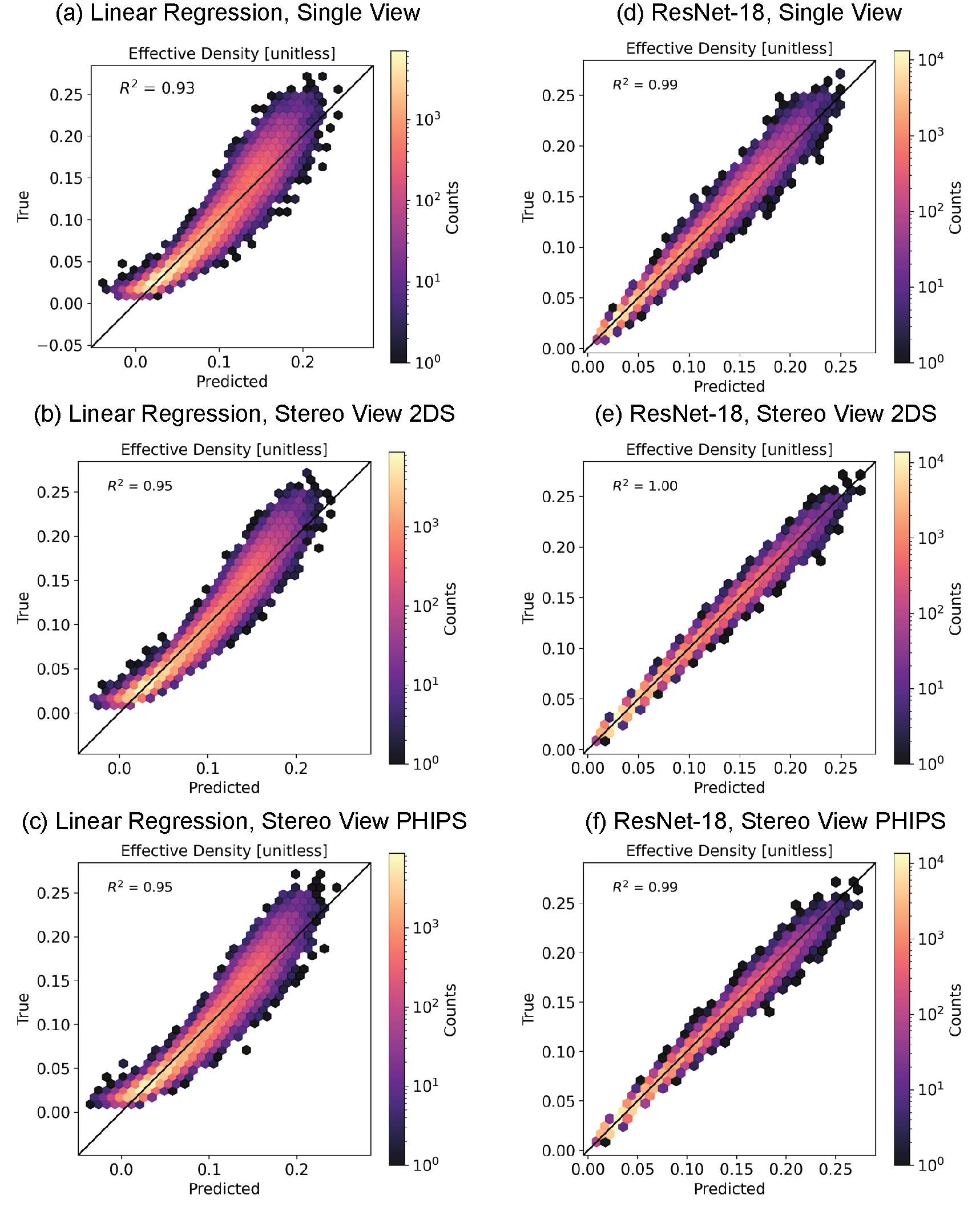}
    \caption{True vs. predicted values for $\rho_e$ between single view (Task 1) regression models (first row) and stereo view (Task 2) regression models (last two rows). Figures (a), (b), and (c) compare single view, stereo 2DS, and stereo PHIPS linear regression models. Figures (d), (e), and (f) compare single view, stereo 2DS, and stereo PHIPs ResNet-18 regression models.}
    \label{fig:stereo-reg-rho}
\end{figure}

\begin{figure}
    \centering
    \noindent\includegraphics[width=\textwidth]{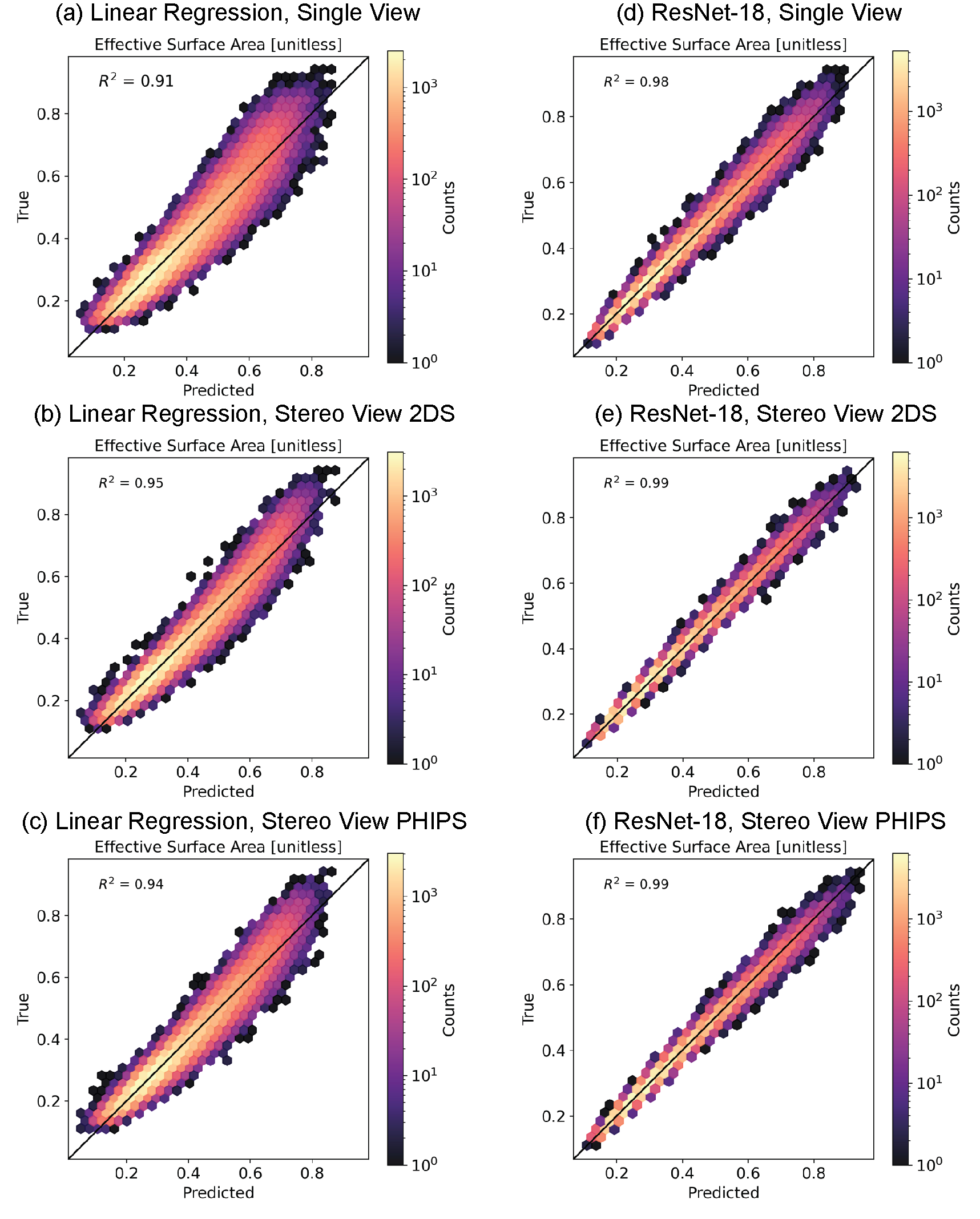}
    \caption{True vs. predicted values for $A_e$ between single view (Task 1) regression models (first row) and stereo view (Task 2) regression models (last two rows). Figures (a), (b), and (c) compare single view, stereo 2DS, and stereo PHIPS linear regression models. Figures (d), (e), and (f) compare single view, stereo 2DS, and stereo PHIPs ResNet-18 regression models.}
    \label{fig:stereo-reg-sa}
\end{figure}

For Task 2 (stereo view), an additional view improved both the linear regression and ResNet-18 models for $\rho_e$ and $A_e$. Scatter plots visually comparing the performance of single view versus stereo view models are shown in Figures \ref{fig:stereo-reg-rho} and \ref{fig:stereo-reg-sa}. We found that the 2DS (90\textdegree\ views) models marginally outperformed the PHIPS (120\textdegree\ views) models, although the performance differences between the two (2DS vs. PHIPS) were smaller than the overall improvements from adding an additional view (single vs. stereo). For linear regression (2DS stereo vs. single view), RMSE and MAE were reduced by 18\% and 9\% for $\rho_e$, and 25\% and 14\% for $A_e$. For ResNet-18 (2DS stereo vs. single view), RMSE and MAE were reduced by 40\% and 23\% for both $\rho_e$ and $A_e$. In general, ResNet-18 models had larger marginal benefits from a second view relative to the linear regression, which highlights the ability of deep learning to more effectively use additional information to improve predictions. Additionally, we note that the positive model bias for both $\rho_e$ and $A_e$ persist for the linear regression models, even with the second view. 

\begin{table}[!htb]
    \centering
    \resizebox{\textwidth}{!}{%
    \begin{threeparttable}[b]
    \caption{Test performance metrics for regression models predicting $\rho_{e}$ and $A_{e}$ for Tasks 1 and 2. RMSE is the root mean squared error. MAE is the mean absolute error. 2DS stereo pairs have 90\textdegree\ viewing angles and PHIPS stereo pairs have 120\textdegree\ viewing angles.}
    \begin{tabular}{l l l l l l l}
    \toprule
    \multirow{2}{*}{Model} &
        \multicolumn{3}{c}{$\rho_{e}$} & \multicolumn{3}{c}{$A_{e}$} \\
    \cmidrule(lr){2-4}\cmidrule(lr){5-7}
        & $R^2$ & RMSE & MAE & $R^2$ & RMSE & MAE\\
    \midrule
    \multicolumn{7}{l}{\textit{Task 1 (Single View) Models}} \\
        LR  & 0.93 & \num{1.17E-02} & \num{9.19E-02} & 0.91 & \num{4.45E-02} & \num{1.85E-01}\\
        RF  & 0.95 & \num{9.68E-03} & \num{8.03E-02} & 0.92 & \num{4.10E-02} & \num{1.76E-01}\\
        MLP  & 0.95 & \num{9.56E-03} & \num{7.96E-02} & 0.92 & \num{4.07E-02} & \num{1.75E-01}\\
        CNN  & 0.97 & \num{7.28E-03} & \num{6.89E-02} & 0.96 & \num{2.85E-02} & \num{1.45E-01}\\
        RN18  & 0.99$^{*}$ & \num{4.75E-03}$^{*}$ & \num{5.48E-02}$^{*}$ 
              & 0.98$^{*}$ & \num{1.86E-02}$^{*}$ & \num{1.15E-01}$^{*}$ \\
    \midrule
    \multicolumn{7}{l}{\textit{Task 2 (Stereo View) Models}} \\
        LR, Stereo 2DS  & 0.95 & \num{9.63E-03} & \num{8.37E-02} & 0.95 & \num{3.32E-02} & \num{1.59E-01}\\
        LR, Stereo PHIPS & 0.95 & \num{9.92E-03} & \num{8.51E-02} & 0.94 & \num{3.50E-02} & \num{1.64E-01}\\
        RN18, Stereo 2DS  & 1.00$^{\dagger}$ & \num{2.84E-03}$^{\dagger}$ & \num{4.22E-02}$^{\dagger}$ 
                          & 0.99$^{\dagger}$ & \num{1.12E-02}$^{\dagger}$ & \num{8.87E-02}$^{\dagger}$ \\
        RN18, Stereo PHIPS  & 0.99 & \num{3.24E-03} & \num{4.46E-02} & 0.99 & \num{1.24E-02} & \num{9.36E-02} \\
    \bottomrule
    \end{tabular}
    \begin{tablenotes}
        \item Note: LR = linear regression, RF = random forest, RN18 = ResNet-18 
        \item $^{*}$ Best Task 1 performance. 
        \item $^{\dagger}$ Best overall performance, including Task 2 models.
    \end{tablenotes}
    \label{table:regression_metrics}
    \end{threeparttable}
    }
\end{table}

\subsection{Predicting number of bullets ($N_b$)}
Classification models for predicting $N_b$ were evaluated using five different metrics: precision, recall, F1 score, balanced accuracy, and top-3 accuracy. In brief, precision is the fraction of positive predictions that were correct, recall is the fraction of true positives that were predicted correctly, F1 score is the harmonic mean between precision and recall, balanced accuracy is the mean of recall and specificity (i.e., true negative rate), and finally, top-3 accuracy is the fraction of correct labels being the top three predictions. Formulas for calculating these metrics can be found in \ref{App:C}. 

For Task 1 (single view), ResNet-18 performed the best for predicting $N_b$, and logistic regression performed the worst. Logistic regression resulted in a precision, recall, F1 score, and balanced accuracy of 0.45, and a top-3 accuracy of 0.90. ResNet-18 resulted in a precision, recall, F1 score, and balanced accuracy of 0.91, and a top-3 accuracy of 1.00. All non-linear classification models performed better than logistic regression, although the random forest and MLP only marginally outperformed logistic regression for some metrics. Notably, the CNN resulted in larger relative improvements over the traditional ML methods for the prediction of $N_b$, compared to the analogous improvements in predictions of $\rho_e$ and $A_e$; suggesting that local spatial correlations in the images are particularly important for the prediction of morphologically-specific features like $N_b$. Table \ref{table:classification_metrics} lists all performance metrics for each Task 1 classification model. 

\begin{figure}
    \centering
    \noindent\includegraphics[width=\textwidth]{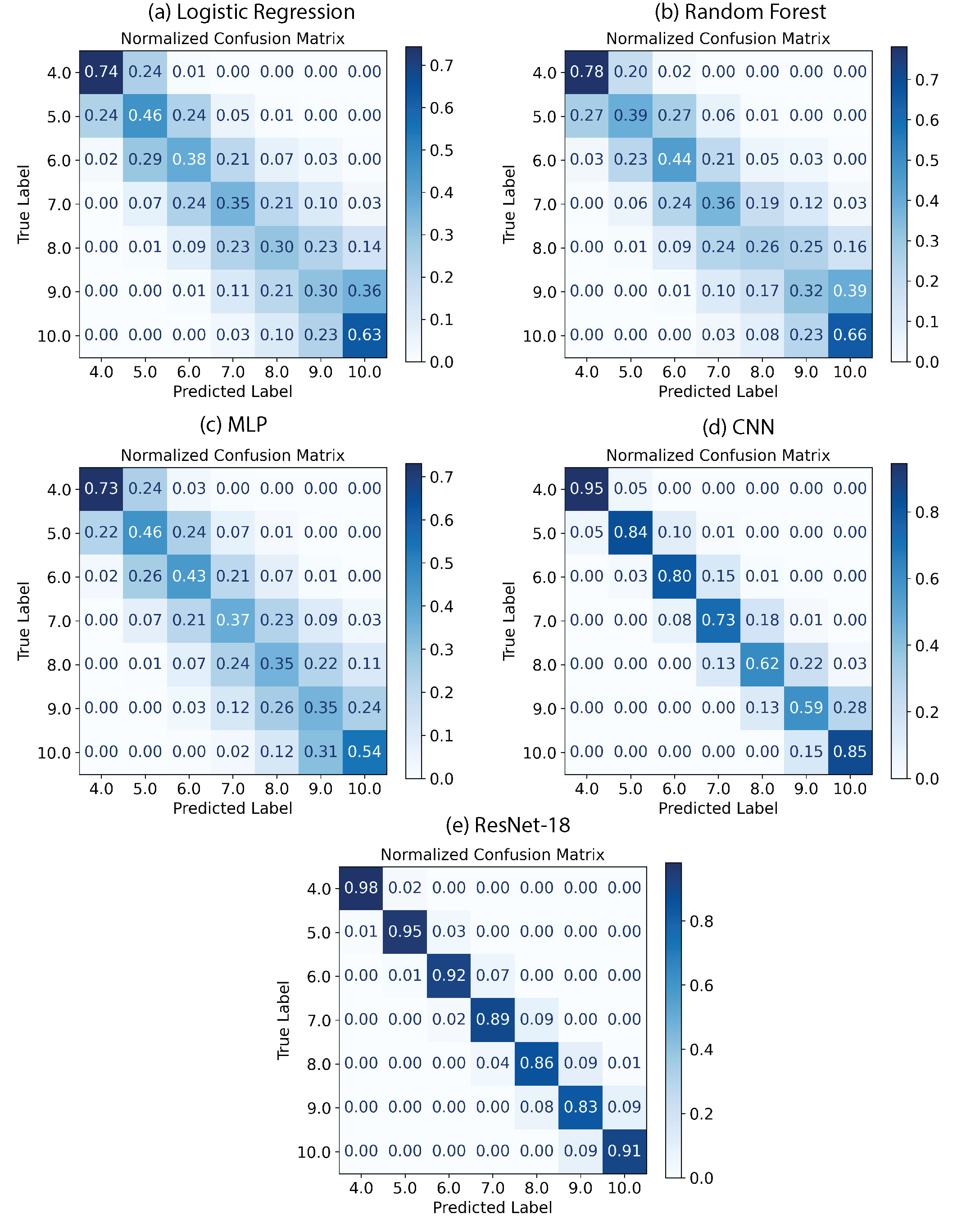}
    \caption{Confusion matrices for each of the five different single view classification models predicting $N_b$.}
    \label{fig:cm-single-view}
\end{figure}

Figure \ref{fig:cm-single-view} shows the confusion matrices for each Task 1 classification model. The value in each cell of the confusion matrix represents the relative frequency of a predicted value (column) given its true value (row). Each row sums to unity, and a perfect model would result in an identity matrix (i.e., 1's in the diagonal and 0's elsewhere). The confusion matrices allow for intuitive comparisons between the different classification models, stratified by class. Reflecting the performance metrics discussed above, the convolutional models (CNN and ResNet-18) were clearly superior in predicting $N_b$. Furthermore, the confusion matrices show how each model performs for each class. For example, we observe that the logistic regression, random forest, and MLP models struggle for $N_b \in \{5, 6, 7, 8, 9\}$. The benefit of introducing convolutional filters in the neural network architecture can be seen by the improvements between Figure \ref{fig:cm-single-view}d and \ref{fig:cm-single-view}c. With our best Task 1 classification model, ResNet-18, we see even greater gains in performance, with the lowest class-conditional accuracy of 0.83 for $N_b = 9$, compared to 0.30 for the logistic regression. 

\begin{figure}
    \centering
    \noindent\includegraphics[width=\textwidth]{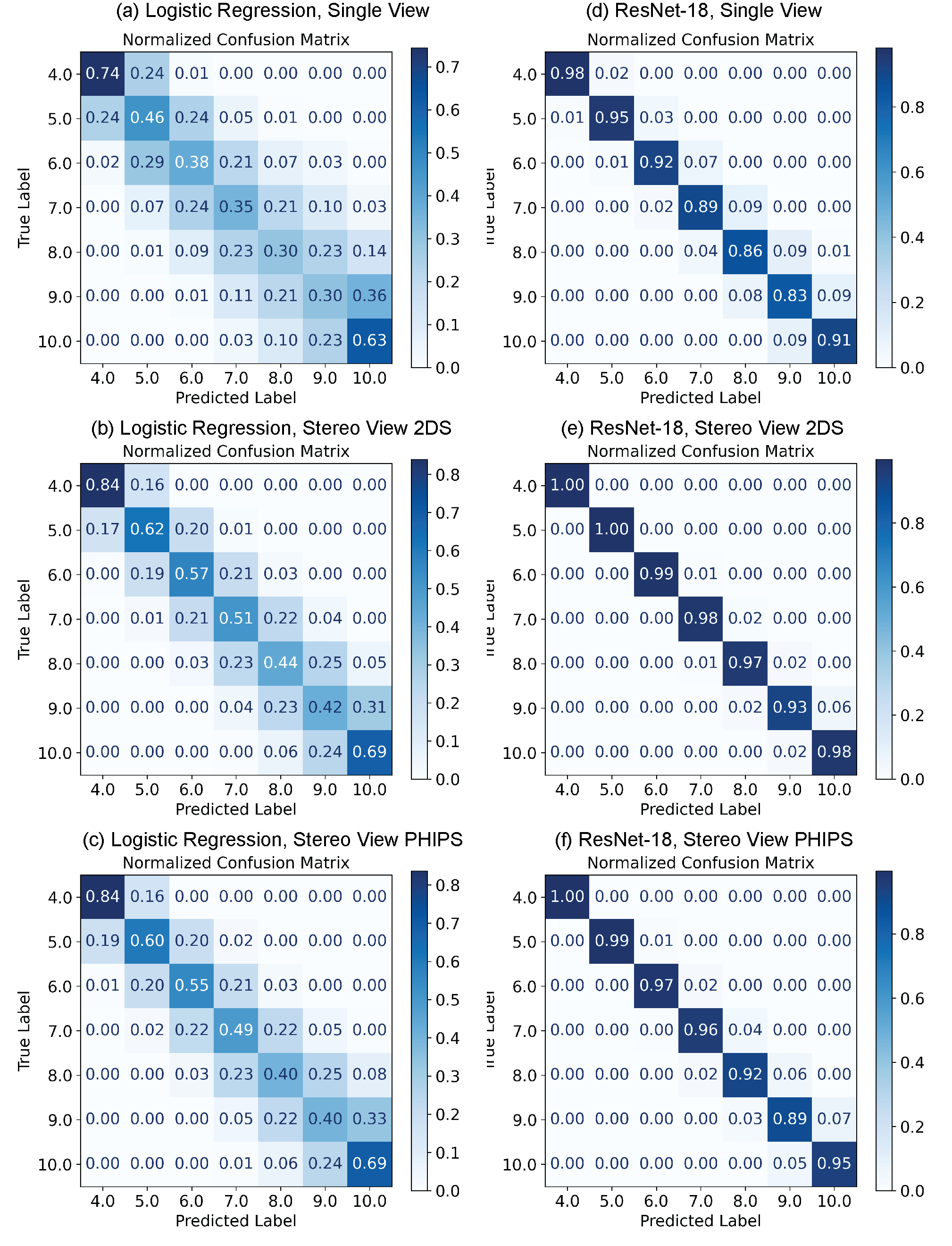}
    \caption{Confusion matrices for the prediction of $N_b$, for logistic regression with single view (a) and stereo views (b) and (c), and for ResNet-18 with single view (d) and stereo views (e) and (f).}
    \label{fig:stereo-cls}
\end{figure}

For Task 2 (stereo view) classification, Figure \ref{fig:stereo-cls} shows the confusion matrices for both single view and stereo view models for logistic regression (left column) and ResNet-18 (right column). Overall, both models benefit from an additional view, but the logistic regression still struggles, particularly for $N_b \in \{5, 6, 7, 8, 9\}$. This is not surprising given the ill-posed nature of this inverse problem. In computer vision, this problem is closely related to classical "shape-from-silhouette" (SFS) reconstruction algorithms, where a 3D shape is estimated from multiple 2D silhouettes (i.e., 2D projections) \cite{baumgartGeometricModelingComputer1974, laurentiniVisualHullConcept1994}. Intuitively, more views generally lead to improved 3D shape predictions, with diminishing returns after a certain number of views. A minimum of two views is needed for classical SFS algorithms, but generally more than two views is recommended for higher-fidelity reconstructions, especially with uncalibrated camera views and no shape priors \cite{buttImportanceSilhouetteOptimization2020, haroShapeSilhouetteConsensus2012, kleinkortVisualHullMethod2017}. Analogously, we find that two views are insufficient to accurately predict $N_b$ with a naive logistic regression approach. 

Using the F1 score improvements as a point of reference, the logistic regression actually benefits more from an additional view (29\% improvement) compared to ResNet-18 (8\% improvement). However, we note that the single view ResNet-18 already had relatively strong predictive performance, with an F1 score of 0.91. For Task 2 classification of $N_b$, ResNet-18 with the 2DS stereo view had the best performance, with a precision, recall, F1 score, and balanced accuracy of 0.98, and a top-3 accuracy of 1.0. Figure \ref{fig:stereo-cls}e also highlights the superior performance of this model, with all elements in the confusion matrix diagonal meeting or exceeding 0.93. 

Overall, we find that $N_b$ can be predicted accurately with ResNet-18, even with a single view. With an additional stereo view, ResNet-18 improves further, with an F1 score of 0.95 and 0.98 for the PHIPS and 2DS versions respectively. We attribute the exceptional performance of ResNet-18 to its deeper architecture (i.e., more hidden layers) which gives the model more expressivity, and its convolutional filters which help the model learn spatial correlations that are important for predicting morphological features such as $N_b$. From an application point of view, our results demonstrate that accurate predictions of $N_b$ are possible for rosettes using only a single view CPI, and additional views (e.g., PHIPS, 2DS) can further boost prediction capabilities when available. 

\begin{table}[!htb]
    \centering
    \begin{threeparttable}[b]
    \caption{Test performance metrics for classification models predicting $N_b$. Top-3 Accuracy refers to the proportion of samples for which the true class was among the three most confident predictions. 2DS stereo pairs have 90\textdegree\ viewing angles and PHIPS stereo pairs have 120\textdegree\ viewing angles.}
    \begin{tabular}{l l l l l l}
    \toprule
        Model &
        \makecell{Precision} &
        \makecell{Recall} &
        \makecell{F1\\Score} &
        \makecell{Balanced\\Accuracy} &
        \makecell{Top-3\\Accuracy} \\
    \midrule
    \multicolumn{6}{l}{\textit{Task 1 (Single View) Models}} \\
        LR  & 0.45 & 0.45 & 0.45 & 0.45 & 0.90\\
        RF  & 0.45 & 0.46 & 0.45 & 0.46 & 0.90\\
        MLP  & 0.46 & 0.47 & 0.46 & 0.47 & 0.91\\
        CNN  & 0.77 & 0.76 & 0.76 & 0.76 & 1.00\\
        RN18  & 0.91$^{*}$ & 0.91$^{*}$ & 0.91$^{*}$ & 0.91$^{*}$ & 1.00$^{*}$ \\
    \midrule
    \multicolumn{6}{l}{\textit{Task 2 (Stereo View) Models}} \\
        LR, Stereo 2DS  & 0.58 & 0.58 & 0.58 & 0.58 & 0.97\\
        LR, Stereo PHIPS  & 0.56 & 0.57 & 0.57 & 0.57 & 0.97\\
        RN18, Stereo 2DS  & 0.98$^{\dagger}$ & 0.98$^{\dagger}$ & 0.98$^{\dagger}$ & 0.98$^{\dagger}$ & 1.00$^{\dagger}$ \\
        RN18, Stereo PHIPS  & 0.95 & 0.95 & 0.95 & 0.95 & 1.00 \\
    \bottomrule
    \end{tabular}
    \begin{tablenotes}
        \item Note: LR = logistic regression, RF = random forest, RN18 = ResNet-18.
        \item $^{*}$ Best Task 1 performance.
        \item $^{\dagger}$ Best overall performance, including Task 2 models.
    \end{tablenotes}
    \label{table:classification_metrics}
    \end{threeparttable}
\end{table}

\subsection{Feature importance using SHAP}
A feature importance analysis was conducted using the SHAP algorithm \cite{lundbergUnifiedApproachInterpreting2017} using the trained MLP models, to probe which features are most relevant to the prediction of $\rho_{e}$, $A_{e}$, and $N_b$. For this analysis, we calculated SHAP values in a \textit{post hoc} fashion for the MLP regression and classification models. For computational efficiency, a subset of 500 test data samples were used for the SHAP analysis, with 100 background samples (also from the test dataset). In brief, SHAP is a popular explainable AI (XAI) method based on game theory that quantifies the contribution of input features on the model output. For more details, we refer readers to \citeA{lundbergUnifiedApproachInterpreting2017}, \citeA{strumbeljExplainingPredictionModels2014}, and \citeA{lipovetskyAnalysisRegressionGame2001}. 

\begin{figure}
    \centering
    \noindent\includegraphics[width=\textwidth]{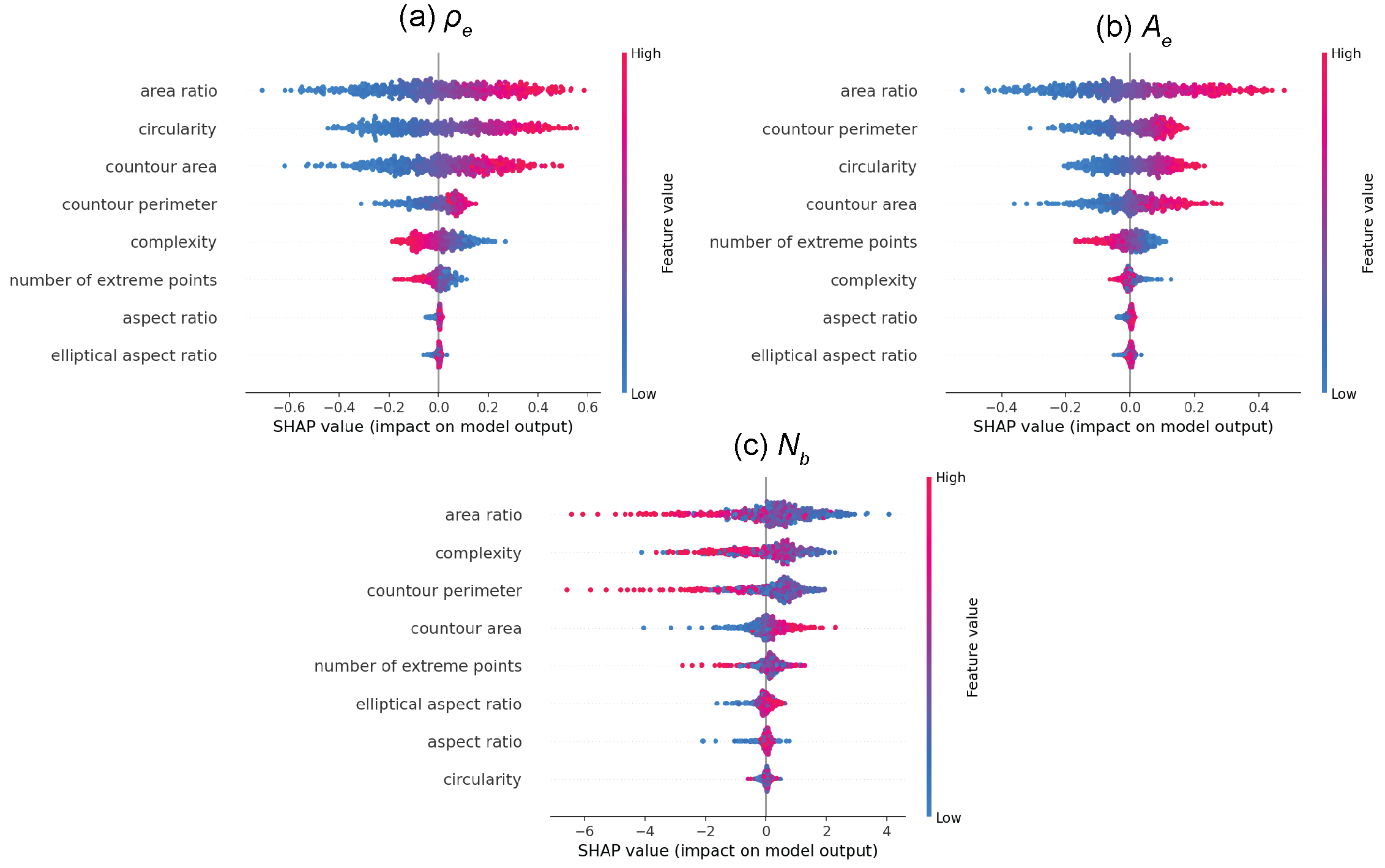}
    \caption{Beeswarm plots showing the SHAP values for (a) $\rho_{e}$, (b) $A_{e}$, and (c) $N_b$. A subset of 500 samples from the test dataset is shown in this figure and the SHAP values shown here are for the MLP models. The color indicates the feature values, the SHAP value magnitude indicates the features impact on the model output, and the sign of the SHAP value indicates the directionality of impact on the model impact.}
    \label{fig:shap}
\end{figure}

Figure \ref{fig:shap} visualizes the feature importance ranking for $\rho_{e}$, $A_{e}$, and $N_b$ in the form of beeswarm plots. Each point on the beeswarm plot is a SHAP value, where larger magnitudes indicate larger influence on the output, and the sign of the value indicates whether the feature impacts the output negatively or positively. The color of each dot displays the value of the feature. We observe that the top few features can explain most of the variance of the model outputs. For $\rho_{e}$, the top three features were area ratio, circularity, and contour area. For $A_{e}$, the top three features were area ratio, contour perimeter, and circularity. And for $N_b$, the top three features were area ratio, complexity, and contour perimeter. 

The SHAP analysis gives some insight into ``black box" nature of the neural network, and at the very least, allows for qualitative confirmation that the important features match intuition. For example, for all predictands, we find that area ratio (i.e., the ratio between the area of the projection and area of the circumscribing circle) is the dominant explanatory feature. This matches physical intuition, especially when examining the directionality of feature values (i.e., the color of points in Figure \ref{fig:shap}) for area ratio. Figures \ref{fig:shap}a and \ref{fig:shap}b indicate that higher values of area ratio correspond to higher values of $\rho_e$ and $A_e$, and vice versa. Figure \ref{fig:shap}c indicates that lower values of area ratio correspond to higher $N_b$ and vice versa. Geometrically, these patterns are logically coherent, and they give some insight into how the MLP is making its predictions. 

One interesting observation from the SHAP analysis is that contour area is present in the top three for $\rho_e$, while contour perimeter is present in the top three for $A_e$. This feature importance ranking is consistent with geometric intuition, given that area is the lower dimensional analog of volume, which is needed to calculate $\rho_e$, and perimeter is the lower dimensional analog of surface area, which is needed to calculate $A_e$. The high ranking of these features is further reassurance that the model is learning sensible relationships between 2D geometric inputs and desired 3D microphysical outputs.  

An important caveat is that the MLP regressor had a moderately good performance for predicting $\rho_e$ and $A_e$ ($R^2 > 0.9$), while the MLP classifier had a relatively poor performance for predicting $N_b$ (F1 score of 0.46). This explains why feature values (point colors) in Figure \ref{fig:shap} exhibit relatively smoother transitions from left to right for $\rho_e$ and $A_e$, compared to $N_b$. Nevertheless, since the MLP classifier performs better than random chance, it exhibits some predictive capability, and these SHAP values can still provide some broad insights into what geometric features may be important for predicting $N_b$.

\section{Application to ice microphysical parameterizations}
Predictions of $\rho_{e}$, $A_e$, and $N_b$ of ice crystals from in situ imagery will allow us to: (1) constrain existing functional relationships in microphysics schemes and (2) establish new functional relationships for the next generation of microphysical parameterizations. 

The most direct application of this framework to existing parameterizations is to use the ML models presented here to predict ice crystal mass and subsequently develop an updated mass dimensional ($m$-$D$) relationship derived from millions of CPI images across multiple field campaigns. Mass-dimensional, and density-dimensional, relationships are widely used to model the mass and radiative properties of ice crystals in numerical weather and climate models. These relationships are often derived from 2D projection images of crystals taken from in-situ observations and, at times, are also derived from geometric models \cite{fridlindDerivationPhysicalOptical2016}. Our method helps extend these prior approaches by connecting these relationships to 3D shapes using ML methods. As mentioned previously, our models are trained to predict $\rho_{e}$ for purposes of being size invariant, but the mass can be estimated as $m = \rho_{e} \cdot V_s \cdot \rho_{ice}$, where $V_s$ is the volume of the enclosing sphere defined as $V_s = \frac{1}{6}\pi D^3$, $\rho_{ice}$ is the assumed density of ice, and $D$ is the maximum dimension length scale that can be estimated from 2D projections.

Our method will complement and build on existing strategies for utilizing in situ measurements to constrain $m$-$D$ relationships. Generally, the $m$-$D$ relationship has been commonly represented by a generic power law relationship, $m = a \cdot D^b$, where $m$ is the particle mass, $D$ is a characteristic length scale (e.g., maximum dimension), and $a$ and $b$ are empirically constrained parameters. Early work used ground based measurements to estimate the mass by melting individual crystals and using accompanying crystal photographs taken before melting \cite{mitchellMassdimensionalRelationshipsIce1990, locatelliFallSpeedsMasses1974, kajikawaMeasurementFallingVelocity1972}. Since then, various studies have attempted to confirm and constrain the $m$-$D$ power law with in situ aircraft measurements \cite{schmittDimensionalCharacteristicsIce2010, leroyIceCrystalSizes2016, erfaniDevelopingBoundingIce2016, jacksonDependenceIceMicrophysics2012}. From a modeling perspective, it is desirable to have a single $m$-$D$ relationship that generalizes for all ice particles, regardless of habit type or size range. However, multiple studies have noted that there is a strong size and habit dependence for the $m$-$D$ relationship, and a single "correct" set of $(a, b)$ values does not exist \cite{schmittDimensionalCharacteristicsIce2010, erfaniDevelopingBoundingIce2016}. 

Past work incorporating in situ optical imagery have often relied on time-integrated, population-level properties, such as ice water content (IWC) and reflectivity ($Z_e$), to validate estimates of $a$ and $b$. The calculation of these bulk properties requires an integration over an assumed size distribution. A direct estimation of mass from single crystal imagery would allow for a bottom-up construction of a $m$-$D$ relationship, along with associated uncertainties. While our method for predicting mass does not directly address the structural deficiencies of using a single $m$-$D$ functional relationship, it opens the possibility of using a bottom-up approach for implicit $m$-$D$ constraints, especially as microphysics schemes move away from discrete ice categories towards more flexible frameworks such as the Predicted Particle Properties (P3) scheme \cite{morrisonParameterizationCloudMicrophysics2015a, morrisonParameterizationCloudMicrophysics2015, milbrandtParameterizationCloudMicrophysics2016}. We also highlight that by constraining geometric parameters from actual observations from the ICEBall campaign, we take a first step towards observationally-driven bottom-up constraints, although we acknowledge the limitations of using limited data from a single field campaign and the need for broader habit representation. 

Additionally, the prediction of other properties will allow for the development of novel  microphysical parameterizations. For example, ice crystal surface area is not commonly used for process level parameterizations, but physical intuition suggests that surface area may be a relevant variable to include in the development of certain process parameterizations, such as vapor depositional growth, terminal fall speed, and radiative absorption and emission. Since our framework allows for the prediction of surface area from imagery, this opens up a possibility of incorporating a semi-empirical relationship between size and surface area into future parameterizations. Similarly, other properties reflecting the geometric and surface complexity of ice crystals (e.g., $N_b$) can be incorporated in ways that would not be possible without some form of empirical or theoretical constraints. Furthermore, predicting $N_b$ from in situ measurements can be important since there is currently no good model for predicting such morphological features from first principles. Accurate predictions of $N_b$ (and other morphological features) from imagery may unlock previously unknown insights into how ice crystals form in the atmosphere and what factors influence their microphysical evolution.

\section{Conclusion} \label{conclusion}

In this study, we present a novel ML framework to predict important 3D microphysical properties of ice crystals from 2D imagery. Using computationally-generated synthetic ice crystals as training data, we developed supervised ML models to predict effective density, $\rho_e$ (proxy for mass); effective surface area, $A_e$ (proxy for surface area); and the number of rosette bullets per crystal, $N_b$; given a single 2D crystal image. We found that our ML models were able to predict $\rho_e$ and $A_e$ skillfully, with best $R^2$ values of 0.99 and 0.98, respectively. Our best single view classification model was able to predict $N_b$ with an F1 score of 0.91 and a top-3 accuracy of 1.0. For all predictands, ML models outperformed linear baselines, confirming that our ML models were able to capture complex, non-linear dependencies between inputs and outputs. Furthermore, we also demonstrated that an additional view can enhance the predictive capability of the ML models for all predictands, highlighting the benefits of stereo view imaging probes such as 2D-S or PHIPS instruments.

Within the broader context of microphysics, our framework provides a path to constrain existing and future parameterizations implicitly with in situ optical imagery. Past studies strongly confirm that representations of ice complexity (i.e., shape and other geometric characteristics) have large impacts on microphysical tendencies and subsequently, weather and climate. By predicting relevant microphysical properties directly from observations, our method allows for bottom-up constraints on relationships such as $m$-$D$ power law parameterizations that will complement existing top-down constraints. In addition to mass, the ability to infer other relevant properties (e.g., surface area) from in situ observations opens the door for new parameterizations that can incorporate information about ice crystal complexity that has not been commonly used in typical schemes. In addition to parameterization development, our framework of using synthetic data for in situ observations can help guide the development and planning of future in situ measurements, similar to how observing system simulation experiments (OSSEs) have been used to quantify the potential benefits of new or additional remote sensing measurements \cite{cucurullObservingSystemSimulation2024}.

Although our methodology shows promise based on test results, there are clear limitations to our study. First, future work must be done to ensure that the synthetic data is actually representative of real ice crystals. This task was out of scope for this current study since our focus here was to demonstrate a functioning proof-of-concept. However, a rigorous and quantitative optimization of the synthetic data is needed in order to fine-tune the ML models to extrapolate well on real CPI images, which are prone to challenges such as noisiness, out-of-focus imagery, and resolution limitations. Additionally, real ice crystals exhibit a lot more complexity than the idealized synthetic models presented in this work. For example, fine-scale morphological details, such as hollowing of bullet arms, are not represented in our rosette model, but should be incorporated in future studies to enhance the applicability of our models to real ice crystals. Second, this framework must be extended to include other ice habits if we want to capture the diversity of all crystal types in ice-containing clouds. This will require the generation of a more diverse synthetic dataset that is representative of the other major habit classes, in addition to bullet rosettes. Finally, our work here was inspired by the desire to make best use of optical imagers such as the CPI, but this inherently limits the scope to larger ice crystals ($\sim$10 to 100 \unit{\um}). Future work can extend an analogous framework to make use of instruments that can measure small ice crystals, such as the Small Ice Detector, which takes in situ measurements of the forwards-scattering signal of particles within the range of about 2 to 60 \unit{\um} \cite{cottonEffectiveDensitySmall2013}.

Our work here demonstrates that it is possible to use computational methods and machine learning to infer crucial microphysical properties of ice crystals from in situ optical imagery. In follow-up studies, we plan to extend this work and ensure that our ML models are robust to real CPI data. The foundational work established in this study will allow for improved observationally-constrained microphysical relationships, and paves the way for future novel parameterizations that better incorporate ice complexity.



\appendix
\section{Description of input features} \label{App:A}
\begin{table}[hbt!]
    \caption{Description of input features used for non-convolutional models.}
    \centering
    \begin{tabular}{l p{8cm}}
    \hline
     Feature Name  & Description   \\
    \hline
      Aspect ratio  &  Aspect ratio calculated from a rectangle ($\leq 1$)\\
      Elliptical aspect ratio & Aspect ratio calculated from ellipse ($\leq 1$)\\
      Number of extreme points & Proxy for how separated the outer most points are on the largest contour \\
      Contour area & Area of the largest contour \\
      Area ratio & Contour area divided by area of encompassing circle \\
      Complexity & Measure of particle intricacy based on \citeA{schmittObservationalQuantificationSeparation2014} \\ 
      Circularity & $4\pi A /P^2$, where $A=area$ and $P=perimeter$ \\
    \hline
    \end{tabular}
    \label{table:input-features}
\end{table}

\section{Neural Network Details} \label{App:B}
\noindent The figures below outline the architectures of all neural networks used in this study.

\begin{figure}[hbt!]
    \noindent\includegraphics[width=\textwidth]{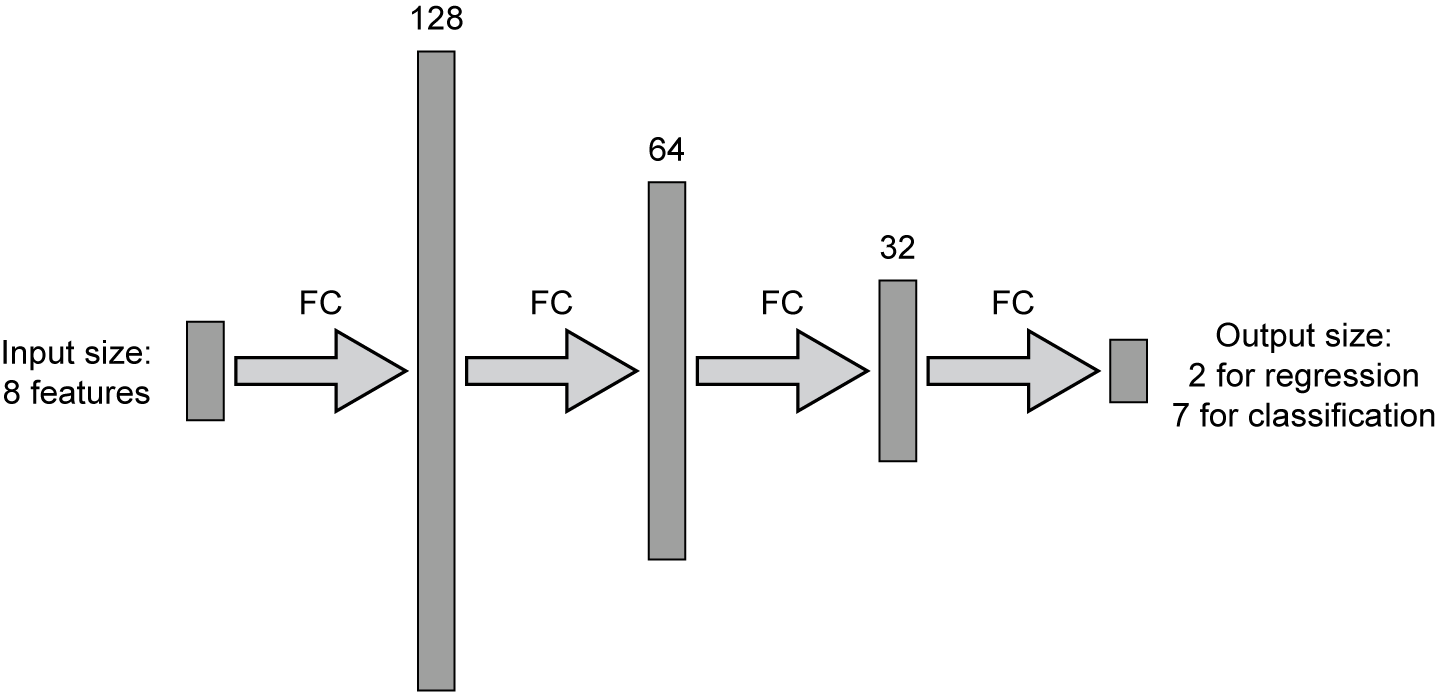}
    \caption{Architecture of the MLPs used in this study. FC stands for fully-connected and the number above each hidden layer signifies the number of nodes in that FC layer.}
    \label{fig:AppC-mlp}
\end{figure}

\begin{figure}[hbt!]
    \noindent\includegraphics[width=\textwidth]{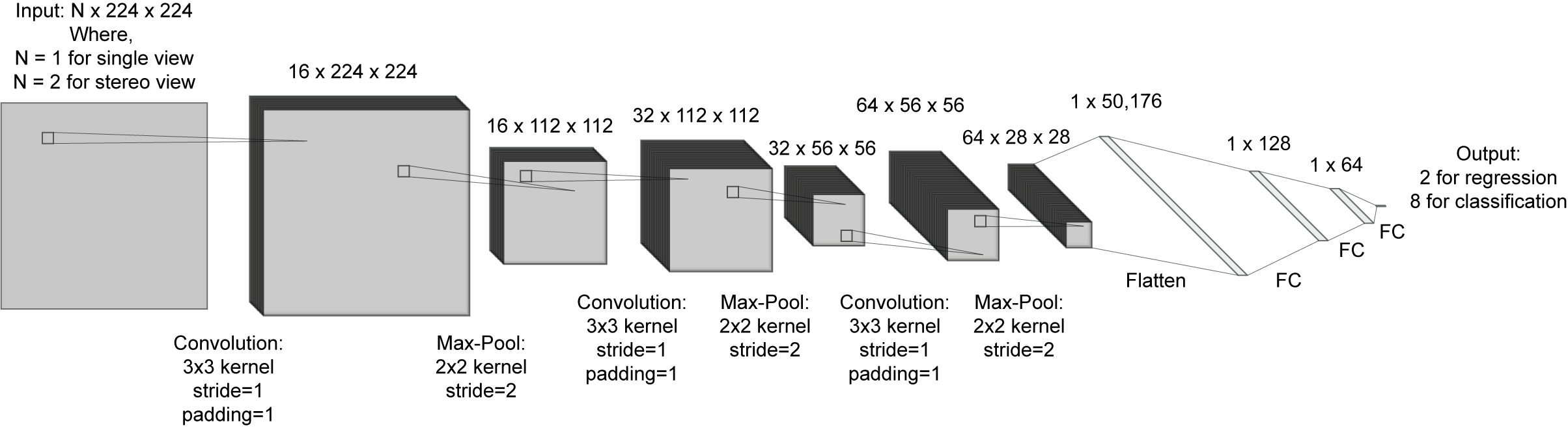}
    \caption{Architecture of the CNNs used in this study. The input had either one or two channels, depending on whether the model was using a single view or stereo view of the crystals. Note: blocks are not proportional to actual dimensions, and are just for illustration.}
    \label{fig:AppC-cnn}
\end{figure}

\begin{figure}[hbt!]
    \noindent\includegraphics[width=\textwidth]{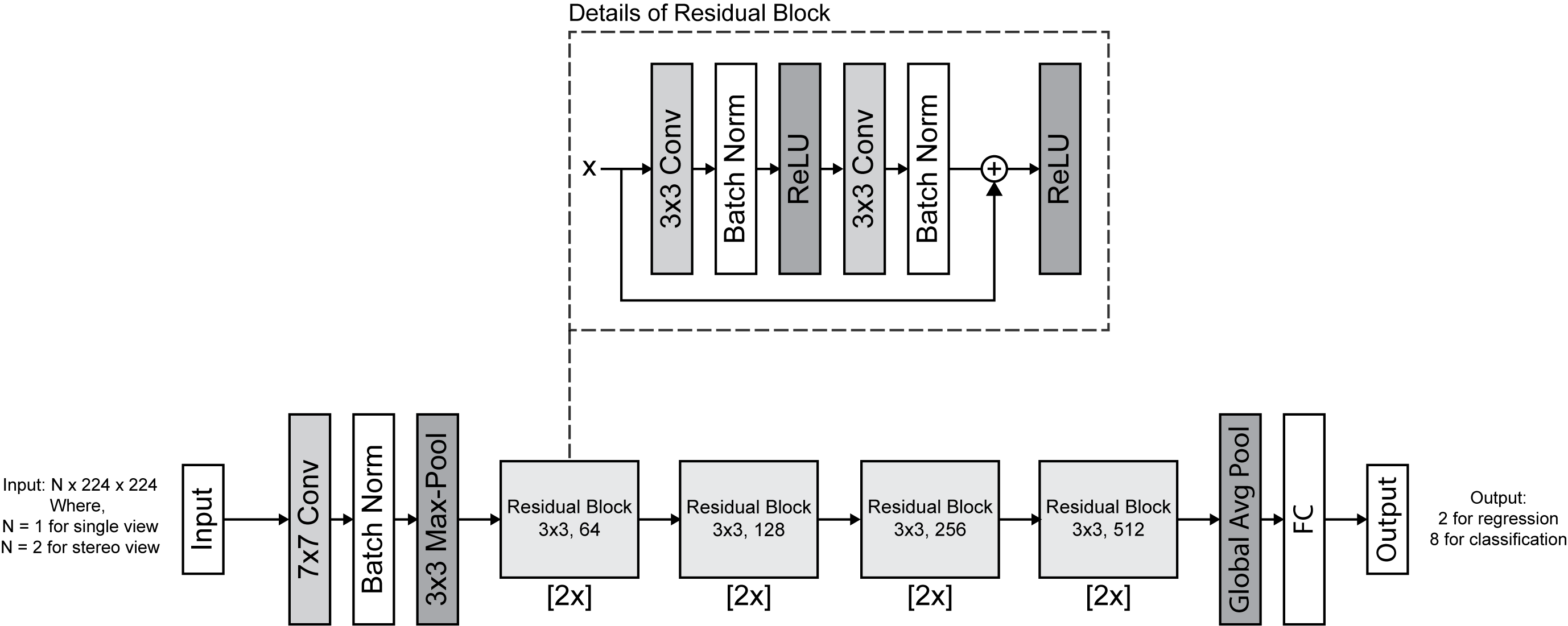}
    \caption{Architecture of the ResNet-18 models used in this study. The input had either one or two channels, depending on whether the model was using a single view or stereo view of the crystals. The inset in dotted lines above the network shows the details of each residual block. The [2x] shown below each residual block indicates that it is repeated twice. Further details regarding ResNet and its architecture can be found in \protect\citeA{heDeepResidualLearning2015}.}
    \label{fig:AppC-resnet}
\end{figure}

\section{Description of performance metrics} \label{App:C}

\noindent Regression performance metrics:

\begin{equation}
R^2 = 1 - \frac{\sum_{i=1}^n (y_i - \hat{y}_i)^2}{\sum_{i=1}^n (y_i - \bar{y})^2}
\end{equation}

\begin{equation}
\text{RMSE} = \sqrt{\frac{1}{n} \sum_{i=1}^n (y_i - \hat{y}_i)^2}
\end{equation}

\begin{equation}
\text{MAE} = \frac{1}{n} \sum_{i=1}^n |y_i - \hat{y}_i|
\end{equation}

\begin{equation}
\text{Precision} = \frac{\text{TP}}{\text{TP} + \text{FP}}
\end{equation}

\noindent Classification performance metrics:

\begin{equation}
\text{Recall} = \frac{\text{TP}}{\text{TP} + \text{FN}}
\end{equation}

\begin{equation}
\text{F1-score} = 2 \cdot \frac{\text{Precision} \cdot \text{Recall}}{\text{Precision} + \text{Recall}}
\end{equation}

\begin{equation}
\text{Balanced Accuracy} = \frac{1}{2} \left( \frac{\text{TP}}{\text{TP} + \text{FN}} + \frac{\text{TN}}{\text{TN} + \text{FP}} \right)
\end{equation}

\begin{equation}
\text{top-}k\ \text{accuracy}(y, \hat{f}) = \frac{1}{n_{\text{samples}}} \sum_{i=0}^{n_{\text{samples}} - 1} \sum_{j=1}^{k} \mathbf{1}\left( \hat{f}_{i,j} = y_i \right)
\tag{C8}
\end{equation}

\noindent
Where:
\begin{itemize}
  \item $n_{\text{samples}}$ is the total number of samples.
  \item $k$ is the number of top predictions considered (i.e., the number of guesses allowed).
  \item $y_i$ is the true class label for the $i$-th sample.
  \item $\hat{f}_{i,j}$ is the $j$-th top predicted class label for the $i$-th sample (ranked by confidence).
  \item $\mathbf{1}(\cdot)$ is the indicator function, which returns 1 if the condition is true and 0 otherwise.
\end{itemize}

\clearpage

%
%

\section*{Open Research Section}
\noindent The synthetic ice crystal dataset (both 2D projections and 3D models) and trained models can be accessed here: 10.5281/zenodo.15758769 \\ 
The code to train and evaluate models can be found here: https://github.com/josephko91/ice3d-ml-paper\\

\acknowledgments
The authors would like to acknowledge funding from the NSF Center for Learning the Earth with Artificial Intelligence and Physics (LEAP) (Award \#2019625) and the Department of Energy (DOE) Atmospheric Science Research (ASR) grants DE-SC0023020, DE-SC00138221, and DE-SC0021033. The authors thank the ASRC Extreme Collaboration, Innovation, and Technology (xCITE) Laboratory for development support, especially Arnoldas Kurbanovas for technical HPC support.

\bibliography{ice_3d_prediction_v3}

\end{document}